\documentclass[twocolumn,pra,aps,superscriptaddress,showpacs]{revtex4}
\usepackage{xspace,amsmath,amsfonts,amsthm,amssymb,amsbsy,graphics,color,graphicx,epsfig}
\usepackage{epstopdf}
\usepackage[breaklinks=true]{hyperref}

\newcommand\foreign[1]{{\it #1\spacefactor=1000}}

\newcommand\ie{\foreign{i.e.}}
\newcommand\etal{\foreign{et al. }}

\newcommand{\sq}[1]{\left[ {#1} \right]}
\newcommand{\tr}[1]{{\textrm {Tr}}\sq{#1}}
\newcommand{\smallfrac}[2]{\mbox{$\frac{#1}{#2}$}}
\newcommand{\half}{\smallfrac{1}{2}}

\newcommand{\ket}[1]{|{#1}\rangle}
\newcommand{\ip}[2]{\langle{#1}|{#2}\rangle}

\newcommand{\expt}[1]{\langle{#1}\rangle}
\newcommand{\dg}{^\dagger}
\newcommand{\D}[1]{{\cal D}\sq{#1}}

\newcommand{\Hc}[1]{{\cal H}\sq{#1}}
\newcommand{\Hcl}[1]{\bar{{\cal H}}\sq{#1}}
\newcommand{\Lc}[1]{{\cal L}_c\sq{#1}}
\newcommand{\Lcun}[1]{{\bar{\cal L}}_c\sq{#1}}

\newcommand{\beq}{\begin{equation}} 
\newcommand{\eeq}{\end{equation}}
\newcommand{\bqa}{\begin{eqnarray}} 
\newcommand{\eqa}{\end{eqnarray}}
\newcommand{\nn}{\nonumber} 

\newcommand{\erf}[1]{Eq.~(\ref{#1})}
\newcommand{\srf}[1]{Sec.~\ref{#1}}
\newcommand{\frf}[1]{Fig.~\ref{#1}}

\newcommand{\hei}{Heisenberg }

\newcommand{\an}[1]{\left\langle{#1}\right\rangle}

\begin{document}
\newcommand{\red}{\color{red}}
\newcommand{\blue}{\color{blue}}
\newcommand{\blk}{\color{black}}
\definecolor{ngreen}{rgb}{0.2,0.6,0.2}
\newcommand{\grn}{\color{ngreen}}

\newcommand{\hmw}[1]{{\color{ngreen} \bf [[{#1}]]}}

\title{Maximum information gain in weak or continuous measurements of qdits: complementarity is not enough}

\author{Joshua Combes}
\affiliation{ARC Centre for Quantum Computation and Communication Technology, and \\ Centre for Quantum Dynamics, Griffith University,
Brisbane, 4111, Australia}
\affiliation{Center for Quantum Information and Control, University of New Mexico, Albuquerque, NM 87131-0001, USA}
\author{Howard M. Wiseman}
\affiliation{ARC Centre for Quantum Computation and Communication Technology, and \\ Centre for Quantum Dynamics, Griffith University,
Brisbane, 4111, Australia}

\begin{abstract}
To maximize average information gain for a classical measurement, all outcomes of an observation must be equally likely. The condition of equally likely outcomes may be enforced in quantum theory by ensuring that one's state $\rho$ is maximally different, or complementary, to the measured observable. This requires the ability to perform unitary operations on the state, conditioned on the results of prior measurements. We consider the case of  measurement of a component of angular momentum for a qudit  (a $D$-dimensional system, with $D=2J+1$  where $J$ is the total angular momentum). For weak or continuous-in-time  (i.e. repeated weak) measurements, we show that the complementarity condition  ensures an average improvement, in the rate of purification, of  only 2. However, 
we  show that by choosing the locally optimal control protocol of this type,  one can attain 
the best possible scaling, $O(D^{2})$, for the  average improvement. For this protocol the acquisition of information is nearly deterministic. Finally we  contrast these results with those for complementarity-based protocols in a register of qbits. \\
\end{abstract}

 \date{\today}
\maketitle 

\section{Introduction}\label{intro}
{In the classical world} predictability is associated with intimate knowledge of a system while unpredictability implies surprise. In information theory, the surprisal of an outcome $k$ is defined as $I(k)\equiv-\log(P_{k})$, were $P_{k}$ is the probability of that outcome \cite{Tri61}. The suprisal is an important quantity as it quantifies the amount of information {one learns} from an outcome of a measurement. {When the average surprisal of an experiment is maximized the observer maximizes the extraction of information from the experiment.} 

In quantum theory the relationship between predictability and knowledge is not clear cut. Even when an observer has maximal knowledge about a quantum system (that is, the observer's state of knowledge is pure), measurement outcomes may be unpredictable. This is the case if measurements on the system are performed in a basis which is not the eigenbasis of the state. (The eigenbasis means the basis in which the operator under consideration, e.g. $\rho$, only has entries on the diagonal.) Nevertheless pure quantum states are the most predictable states available. It is for this reason, and others eloquently explained in Ref.~\cite{FucJac01}, that we take the impurity to be a measure of information. These considerations have lead to information \cite{FucJac01} and control \cite{DohJacJun01} theoretic formulations of information--disturbance relations. 

Inspired by this work, Jacobs asked the question: how quickly on average can information be extracted from a quantum system using continuous quantum measurements in an complementary basis \cite{Jac0303}. (Other ways of quantifying the rate of information gain can also be considered~\cite{WisRal06,ComWisJac08}.) He found, for a widely applicable continuous measurement model, that measuring in an complementary basis led to a speed-up in the purification by a factor of two for a two level system \cite{Jac0303}. {(Practically, such a rapid purification protocol can be a primitive for a rapid state preparation protocol \cite{WisRal06,ComWisSco10}.) Later this was generalized to complementary bases for $D$-dimensional systems. In Ref.~\cite{ComJac06} the lower bound on the speed-up for $D$-dimensional systems was found to be $S_{\rm LB}=(2/3)(D+1)$ when the monitoring was performed in an complementary basis. Recently in Ref.~\cite{ComWisJacOCo10} the upper bound $S_{\rm UB}$ was proven to be bounded above by $D^{2}/2$ for measurement in any complementary basis. 

In this paper we give an explicit construction of an complementary measurement protocol for a qudit
(a $D$-dimensional system, with $D= 2J +1$ where $J$ is the total angular momentum) based on the quantum Fourier transform (QFT) which achieves {a speed-up of $0.2D^{2}$. Further we show that transforming the state's eigenbasis so that it is complementary with respect to the measurement eigenbasis is {\em not} sufficient to achieve this speed-up; the speed-up can be as low as $2$.  In addition to this we show, numerically, that unbiased bases can achive the upper bound, saturating the inequality $S \le D^{2}/2$, for $D$ even.} To define these speed-ups rigorously, it is necessary to derive rigorous bounds on information extraction rate for commuting measurements without control, so we also do that in this paper. (For the rate of commuting measurements with control see Refs.~\cite{ComWisJac08,ComWisSco10}). Finally we consider measurements in commuting and complementary bases for a register of qbits. The complementary measurement scheme based on the QFT appears only to give a speed-up of 2 in the case of a register. 
 
\section{further background and structure of this paper}
In this section we explain the structure of our paper by reviewing the relevant prior work on the subject and relating it to the work presented in this article. 

In the quantum information context, information--disturbance relations were inspired by the {\em epistemic} interpretation of quantum states \cite{Spe08} as applied to quantum cryptography \cite{FucJac01}. The idea was that the information gathering and disturbance should be grounded with respect to an observer's state of knowledge, not some pre-existing property of the system. To formalize this intuition Fuchs and Jacobs in Ref.~\cite{FucJac01} considered something like the following game. Two observer's Alice and Bob initially agree that the state of some quantum system is $\rho_{0}$, an impure state. At some later time Charlie will perform a measurement along a randomly chosen axis and ask Alice and Bob to predict the outcome of the measurement. After repeating this many times Charlie rewards the individual who makes the most correct guesses. Alice, being clever and resourceful, has the ability to perform one unitary operation followed by a two-outcome measurement. Her aim is to refine her state of knowledge about the quantum system, thus increasing its predictability and accordingly the likelihood of receiving the money. However, Alice's morals prevent her from intentionally sabotaging Bob for the pecuniary reward. Thus Alice would like to increase the accuracy of her predictions without affecting those of Bob. 

This formulation suggests (intuitively) that weak measurements are the right way to probe the system. From the point of view of the weak measurement formalism, a projective measurement corresponds to a measurement of infinite strength \cite{FucJac01}. For finite strength measurements on a single qbit it was found that there is a non-trivial tradeoff between Alice's information gathering actions (the refinement of her predictability) and the disturbance to Bob's predictability (his state of knowledge). 

For a fixed measurement rate the interval of the non-trivial tradeoff may be characterized by the angle between the Bloch vector and the measurement axis. There are two extremes of this tradeoff. Minimal disturbance and minimal {\em average} information gain occurs when Alice's axis is aligned with the Bloch vector, that is the measurement eigenbasis and the state's eigenbasis commute. Maximal disturbance and maximal average information gain, is attained when Alice's axis is orthogonal to the Bloch vector. This strategy corresponds to making the outcomes equi--likely, that is maximizing the {\em suprisal}~\cite{Tri61} of the measurement result. In this situation the bases are said to be complementary, maximally non-commuting, or unbiased. Thus we conclude that quantum theory does not allow Alice to succeed. The ability to measure in different bases is an essential feature of quantum mechanics which provides the richness to the problem considered here. 


It is easy to understand how non-commuting measurements provide maximal information gain and maximal disturbance through the following picture. Consider a mixed qbit state with its Bloch vector aligned along the $+\ket{z}$ direction. By measuring in the $\ket{x}$ basis the Bloch vector will elongate regardless of the measurement outcome. This point is illustrated in \frf{fig0} (a). Furthermore both outcomes are equally likely, which means the {\em average} information extraction is maximal. Unfortunately the post-measurement state is not pointing in the same direction as the pre-measurement state, so it will not commute with the original state. Conversely if the measurement basis and the state are along the $\ket{z}$ direction then, due to the stochastic nature of the measurement outcomes, the average elongation of the Bloch vector will be minimal \cite{WisRal06,WisBou08}. In any particular measurement the impurity may increase or decrease. But the disturbance to the state will be minimal \cite{FucJac01}. In \frf{fig0} (b) we illustrate that one of the possible measurement outcomes causes the Bloch vector to shrink whist the other elongates it.

\begin{figure}[h!]
\begin{center}
\includegraphics[width=0.8\hsize]{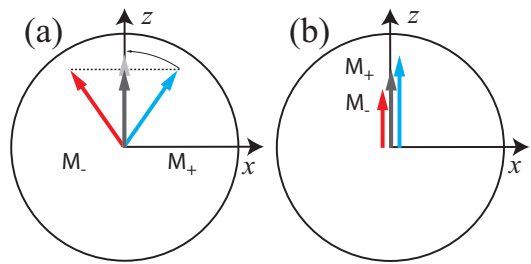}
\caption{A Bloch sphere depiction of the extremes of information--disturbance tradeoffs. In both figures the dark arrow represents the Bloch vector of the pre-measurement state. The arrows labeled with $M_{\pm}$ correspond to the post measurement states conditioned on the random outcome of a weak measurement. Figure (a) is measurement of a state in a complementary or unbiased basis, in this case the measurement axis is along the $x$ direction. When this occurs the Bloch vector always elongates, as indicated by the light grey arrow (the plus outcome rotated for easy visual comparison). Figure (b) depicts the commuting measurement. In this case if the measurement result $M_{-}$ is obtained the Bloch vector decreases in length.}\label{fig0}
\end{center}
\end{figure}

While information--disturbance relations are interesting for single measurements the subject has been thoroughly explored. Less exploration has been undertaken in continuous measurement setting due to the increased difficultly of the analysis. In order to make some progress the focus on disturbance has mostly been ignored (for good reasons as we shall see). That is, the focus has been on information gathering. This restriction has led to an interesting development in the theory of continuous quantum measurements \cite{Car93, ConMeas,WisMil10} . 

To demonstrate an improvement in the extraction of information the extraction rate of a standard continuous measurement must be defined. We follow Jacobs and define the standard measurement as a measurement in a commuting basis \cite{Jac0303}. This is sensible for a number of reasons. Firstly measurement will rapidly destroy the coherences of any state that does not commute with the measured observable \cite{Zur02}. This can be understood as the measurement projecting the state to the eigenbasis of the observable. To counter this effect an adaptive measurement or closed loop feedback control is required. The second reason is a continuous measurement in a commuting eigenbasis is essentially a classical continuous measurement. 

The analysis of disturbance  and  information gain is significantly harder for continuous measurements of $D$-dimensional systems  \cite{StoVanMab04,ComJac06}. Minimal disturbance still occurs when measuring in the same eigenbasis as the state (a commuting basis). A full analysis of this situation would require characterization of the distribution of information extraction rates, allowing calculation of the average information extraction, median, worst case and best case. In \srf{CommMeas} we rigorously derive the average purification rate for a commuting measurement of a $D$-dimensional system and provide an asymptotic expression. Additionally we give exact results for bounds on the spread of the distribution of purities, in \srf{SEC_explain}, which were heuristically derived in previous work \cite{StoVanMab04}. 

Now that we have defined what a standard measurement is we can explore what it means to increase the rate of extraction of information. In the context of control theory one might be interested in revealing or acquiring the value of some parameter; that is the domain of adaptive measurements \cite{WisMil10,adaptme}. But more typical control objectives are to drive the evolution of the state or to stabilize the state against noise \cite{WisMil10,StoVanMab04,WisManWan02,fbme,JacLun07}. Thus disturbing the prior state is not a problem. This why disturbance is not usually considered in feedback control. What becomes important is the noise introduced by the measurement to the system. Doherty, Jacobs and, Jugnman in Ref.~\cite{DohJacJun01} showed, for a continuously monitored qbit, that the measurement noise is maximal when the eigenbases of the state and measurement axis are unbiased. 

The study of Wiseman, Mancini and Wang in Ref.~\cite{WisManWan02}, explored state stabilisation in a qbit. The relevant result from Ref.~\cite{WisManWan02} to the present discussion relates to the difficulty of stabilising a state which is complementary to the measurement basis. They showed that this can be achieved in the limit of infinite strength ``Bayesian'' feedback, but not in simpler ``Markovian'' feedback procedures. This is a consequence of the increased measurement noise as predicted by Doherty \etal~\cite{DohJacJun01}. Later Jacobs showed how continuously monitoring the state of a qbit in a complementary basis (assuming infinite strength feedback) leads to a doubling rate of information extraction over monitoring in a commuting basis asymptotically \cite{Jac0303}. 

Recently a number of studies have examined the use of weak (and continuous) measurement in a complementarity basis in systems of dimension $D$ greater than two. The first study showed that the lower bound on the improvement in the rate of information extraction when measuring in any complementary is $(2/3)(D+1)$ \cite{ComJac06}. The second study showed this {\em speed-up} is upper bounded by a factor $D^{2}/2$ \cite{ComWisJacOCo10}. We summarise these results in \srf{L_Dfb_sec}. In Secs.~\ref{nopermsec} to \ref{discusqft} of this paper we give an explicit construction of an unbiased basis protocol for an arbitrary finite dimensional system based on the quantum Fourier transform. We derive upper and lower bounds on the speed-up of the QFT protocol. Surprisingly the numerics show that the protocol saturates the upper bound -- for even $D$. {For completeness we mention the recent work of Ruskov et al. \cite{RusKorMol10}, where it was shown that monitoring a qbit in three complementary bases where they claimed a speed-up $S\sim 3$. We are currently exploring this relationship so we will not discuss it further in this article.}

Motivated by some of the above investigations Jacobs and Lund 
argued that the task of state stabilisation is best accomplished by measurement in an complementary basis; in the regime of strong feedback \cite{JacLun07}. This is one reason to continue investigations into maximal information extraction, foundational issues aside. In Ref.~\cite{JacLun07} the authors raised two interesting questions: first, are all complementary bases are equally good for information extraction when $D\ge4$? Second, is measurement in a complementary basis is best for information extraction?

We reexamine these questions in \srf{whycomp} and \srf{dl_all_mub}. We show that complementary, i.e. equi-likelihood condition, is not sufficient to guarantee maximum information extraction. Our analysis provides an insight into the mechanism for information extraction, which is quantum mechanical backaction. From this we explain how maximum information extraction is achieved by enforcing the complementary condition and arranging the eigenvalues in phase space so measurement backaction can have maximum effect. We provide further evidence for the claim that not all bases are equally good. Because Jacobs and Lund did not examine permutations of the unbiased basis it was still unclear from their analysis whether all unbiased bases are equally powerful for information extraction. So we also consider the role of permutations in \srf{dl_all_mub}.  Our results show that even after maximizing information extraction over permutations of the unbiased basis, not all unbiased bases are equally good at information extraction.

In \srf{qftreg} we examine complementary measurements in a register of qbits using the QFT. We start our analysis, in \srf{regnfb}, with the case of commuting measurements~\cite{ComWisJacOCo10}. Then, in \srf{SEC_reg_lb}, we review the results of Ref.~\cite{ComWisJacOCo10} where it was shown that the speed-up bounded by $2n(2^{n}-1)^{-1}\le S \le 2n$. In \srf{SEC_QFT_reg} we construct a protocol based on the QFT and present results which suggest the speed-up is  at most $2$. We conclude in \srf{discuss}.


\section{Commuting Measurements}\label{CommMeas}
Later in this article we will investigate strategies that increase the rate at which information can be extracted from quantum systems. Before these strategies are examined, we must fully characterize this rate in the absence of feedback. This will provide a benchmark to assess the performance of the strategies that aim to increase the rate of information extraction.

In this section we restrict our attention to continuous measurement (CM) of a single $D$-dimensional quantum system, also called a qdit, monitored by a single output channel. This output channel is monitored in such a way that it results in a diffusive quantum trajectory~\cite{Car93,WisMil10}. This can be thought of as a sequence of short-duration, weak measurements. Following Fuchs and Jacobs~\cite{FucJac01} our notion of information is characterised by the impurity: $L[\rho(t)] \equiv 1- \tr{\rho(t)}$. Accordingly we derive an exact expression for the average impurity $\expt{L(t)}$ and a simpler approximate expression in \srf{L_nfb_sec_analysis}. In \srf{compareexact} and \srf{compare_stoc} we examine the validity of our approximate expression. To completely characterise the rate of information extraction from commuting measurements we look at bounding the distribution of purities in \srf{SEC_explain}. Then finally we summarise the key messages in \srf{nfbsum}.

\subsection{Qdit - impurity analysis}\label{L_nfb_sec_analysis} 

The starting point for the current analysis is similar to that presented in Refs.~\cite{Jac04,ComJac06, ComWisJacOCo10}. We use a widely applicable model that describes the evolution of the state $\rho$ of a quantum system due to repeated weak measurements of a dimesionless observable $X$. This model is described by the stochastic master equation (SME)~\cite{WisMil93,CMreview2, CMreview1,WisMil10} 
\begin{eqnarray}\label{SME}
 d\rho= dt \Lc{X} \rho \equiv  
  2\gamma \, dt\, \D{X}\rho +\sqrt{2\gamma}\,dW\,\Hc{X}\rho,  
  \label{eq:drho1}
\end{eqnarray} 
where $\D{A} \rho \equiv A\rho A\dg -\half (A\dg A \rho + \rho A\dg A)$, $\Hc{A} \rho \equiv  A\rho +\rho A\dg - \tr{(A\dg+ A )\rho}\rho$ \cite{WisMil10}, and $dW$ is the increment of a Wiener noise process~\cite{Jacobs10}. It should be noted that we have moved to a frame that has enabled us to factor out the Hamiltonian evolution. The {\em measurement} {\em rate}, \blk $\gamma$, determines the rate at which information is extracted, and thus the rate at which the system is projected onto a single eigenstate of $X$~\cite{AdlBroBruHug01, StoVanMab04,VanStoMab0605,vanHStoMab05}. (This means for times $\tau \gg \gamma^{-1}$ we may say that we have performed a projective measurement of the observable $X$.) The measurement  result in a small time interval $[t,t+dt)$ is denoted
\begin{eqnarray}
dR(t) = \sqrt{4\gamma}\langle X(t) \rangle dt + dW(t),
\end{eqnarray}
where $dW$ is the same Wiener noise process that appears in Eq.(\ref{eq:drho1}), and $\expt{X(t)} = \tr{X\rho(t)}$. A quantum trajectory is a continuous--in--time description of the state conditioned on the measurement {result}. {Thus,} to specify a trajectory over the interval $[0,T)$ we must be given the measurement results over that interval. 
Then the It\={o} definition $\rho(t +dt) =\rho(t) +d\rho(t)$ may be used recursively to generate a trajectory. Clearly a single weak measurement is described by one application of this relation.

 We will denote the continuous measurement record obtained by the observer integrated up until time $t$ as 
\begin{equation}
R(t) = \int_0^t dR(t').
\end{equation} 
The unnormalised version of \erf{SME} is know as the {\em linear} SME. We denote this linear SME as \cite{GoeGra94,Wis96,JacKni98,WisMil10}
 \begin{eqnarray}\label{SMEunorm}
 d\bar{\rho}=dt\Lcun{X}\bar{\rho} \equiv   
 2\gamma \, dt\, \D{X}\bar{\rho} +\sqrt{2\gamma}\,dR\,\Hcl{X}\bar\rho .  
 \end{eqnarray}
Here $\Hcl{A} \rho \equiv  A\rho +\rho A\dg$, and the bar over $\rho$ denotes the lack of normalization at all but the initial time. We take the initial state to be $\bar{\rho} = \rho(0) = \textbf{I}/D$ where $\mathbf{I}$ is the $D\times D$ identity matrix. Because the initial state matrix commutes with $X$ at all times this makes obtaining the solution to \erf{SMEunorm} simple. This solution is~\cite{CMreview2} 
\begin{eqnarray}\label{St_rhotildejz}
 \bar{\rho}(R,t)&=&\exp(-4\gamma X^2t)\exp(2\sqrt{2\gamma}XR(t))\textbf{I}/D.
\end{eqnarray}

Using the linear trajectory (see Refs.~\cite{GoeGra94,Wis96,JacKni98}) solution, \ie\ \erf{St_rhotildejz}, we now explicitly calculate the form of the $D$-dimensional state matrix.  We take the observable $X$ to be the $z$ component of angular momentum (represented by the operator  $J_z$), thus $D = 2J+1$. In matrix form the $J_z$ operator is $J_z = \mathrm{diag}(J,J-1,\ldots,-J+1,-J)$. Using this representation of $J_z$ the unnormalized solution of the state evolution can be written as matrix elements exponentiated:

\begin{eqnarray}\label{rhodiag}
\nn \bar{\rho}(R,t)&=&\frac{1}{D}\textrm{diag}\Big (  e^{-4\gamma J^2t}e^{2\sqrt{2\gamma}J R},\\&& \nn\phantom{\frac{1}{D}\textrm{diag}m}  e^{-4\gamma (J-1)^2t}e^{2\sqrt{2\gamma}(J-1) R},\\&& \phantom{\frac{1}{D}\textrm{diag}m}\cdots, e^{-4\gamma (-J)^2t}e^{-2\sqrt{2\gamma}J R} \Big ) .
\end{eqnarray}

The normalization factor can be written with a non-symmetric sum or a symmetric sum
\begin{eqnarray}\label{qdit_norm}
  \mathcal{N}\equiv\tr{\bar{\rho}}
    &=& \frac{1}{D}\sum_{m=0}^{2J}e^{-4\gamma(J-m)^2t}e^{2\sqrt{2\gamma}(J-m)R}\label{normnonsym}\\
    &=& \frac{1}{D}\sum_{s =-J}^{J}e^{-4\gamma s^2t}e^{2\sqrt{2\gamma}sR}.
\end{eqnarray}
For discussions later we also calculate the probability distribution of the result $R$ and choose the symmetric sum for the normalization
\begin{eqnarray}\label{probdisjz}
dR \,\mathcal{P}(R)&=& dR\,\tr{\bar{\rho}(R,t)}P(R )\\
&=&dR\frac{1}{D\sqrt{2 \pi t}}\sum_{s=-J}^{J}e^{-(R-2\sqrt{2\gamma} s t)^2/2t}.
\end{eqnarray}
Here $P(R)$ has been called the ostensible distribution for $R$ \cite{WisMil10,Wis96}. It is given by $P(R,t)= \exp{(-R^{2}/2t)}/\sqrt{2\pi t}$. By choosing the symmetric summation for the normalization one can see that the probability distribution for $R$ contains $D$ peaks centred around the values $\propto s$, where $s\in [J,J-1,\ldots,-J+1,-J]$.  For times $t\gg \gamma ^{-1}$, the distribution is sharply peaked about these $D$ values. Each of these peaks has a FWHM of $2\sqrt{2  \ln 2}\sqrt{t}\approx 2.35 \sqrt{t}$. This seems paradoxical; one would expect that as time increases the distribution would get narrower, reflecting the fact that the observer is more confident about which eigenstate they actually have. This paradox is resolved when one realizes that the distance between the peaks increases at a rate greater than the increase in the width of the peaks. This time-dependent scaling of the distance between the peaks can be removed by changing variables to $V \equiv R/(2\sqrt{2\gamma}t)$. Under this transformation the probability distribution is
\begin{eqnarray}\label{probdisjz_scaled}
dV\mathcal{P}(V)&=&dV\frac{1}{D}\sqrt{\frac{4\gamma t}{\pi}}\sum_{s=-J}^{J}e^{-(V- s)^2/(1/4\gamma t)}.
\end{eqnarray}
Now the FWHM of each peak scales as $\sqrt{\ln{2}/(\gamma t)}\approx 0.83/ \sqrt{\gamma t}$, which clearly illustrates that the probability distribution becomes sharply peaked about the $D$ values for long-times. In \frf{regions} the dashed curve is a plot of \erf{probdisjz_scaled} for $D=5$ and $t=4\gamma^{-1}$. All of the peaks are clearly distinguishable with a measured FWHM $\approx 0.418$. This agrees with the prediction given by $0.83/\sqrt{\gamma t}$.

Returning to the non-symmetric sums, i.e. \erf{normnonsym}, for the remainder of this calculation, the purity is
\begin{eqnarray}
  \tr{\rho^2}&=&
\frac{1}{D^2\mathcal{N}^2}\sum_{m=0}^{2J}e^{-8\gamma(J-m)^2t}e^{16\gamma t(J-m)V}.
\end{eqnarray}
The impurity, for a particular value of $V$, is given by
\begin{eqnarray}
\Lambda(V,t)&=&\bigg(1-\frac{\sum_{m=0}^{2J}e^{-8\gamma(J-m)^2t}e^{16\gamma t(J-m)V}}{D^2\mathcal{N}^2}\bigg)\label{notnum}\\
\nonumber &=&\bigg(\frac{D^2\mathcal{N}^2-\sum_{m=0}^{2J}e^{-8\gamma(J-m)^2t}e^{16\gamma t(J-m)V}}{D^2\mathcal{N}^2}\bigg).\\\label{numericz}
\end{eqnarray}
The solid curve in \frf{regions} is a plot of \erf{numericz} for $D=5$ and $t=4\gamma^{-1}$. To find the average value of impurity for a continuous measurement of $J_z$ one must integrate this function over the scaled measurement record $V$, weighted by $P(V)$~\cite{ComJac06}:
\begin{figure}[h!]
\begin{center}
\leavevmode \includegraphics[width=\hsize]{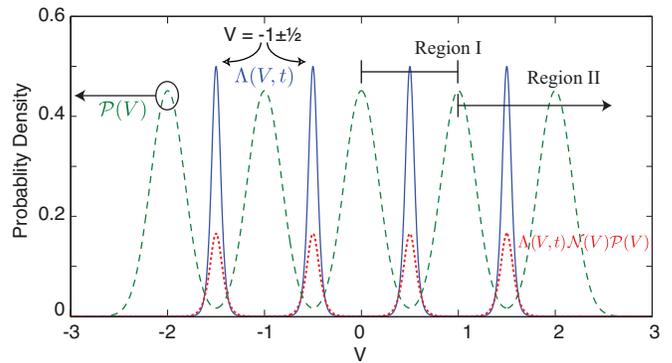}
\caption{The dashed curve is the probability distribution $\mathcal{P}(V)$, \erf{probdisjz_scaled}; the solid line is $\Lambda(V,t)$, \erf{numericz}; shown for $D=5$ at $t=4\gamma^{-1}$. The dotted curve is $\Lambda(V,t)\mathcal{N}(V)P(V)$ multiplied by $20$ (so that it can be viewed on the same scale). Also plotted are regions for the two eigenvalue expression for the impurity integral.} \label{regions}
\end{center}
\end{figure}

\bqa
\expt{L(t)}&=&\int_{-\infty}^{\infty}\Lambda(t,V)\mathcal{N}(V)P(V)dV.\label{impint}
\eqa
When using this result to calculate the measurement-induced evolution of the mean linear entropy one should use the kernel in \erf{numericz} rather than the kernel in \erf{notnum}. This is because it will reduce the numerical error that occurs when subtracting from one a number very close to one.

To calculate this integral one must perform numerical integration. It would be useful to have a fully analytic expression for $\expt{L(t)}$ in the $D$-dimensional case. Now we will make some approximations to obtain a simple analytic expression for \erf{impint}. We will show that the approximations allow one to place a bound on the full expression. Then we will confirm our approximations with numerical simulations in \srf{compareexact} and \srf{compare_stoc}.

The first approximation involves truncating the state matrix to two eigenvalues and then renormalizing. We call this the {\em two eigenvalue approximation}. The motivation for the truncation stems from the following observations. In the long-time limit ($t\gg \gamma^{-1}$) the true probability distribution, $\mathcal{P}(V)=\mathcal{N}(V) P(V)$, is sharply peaked in $D$ places; see \frf{regions}. Also in this limit one finds that $\Lambda(V,t)$, \erf{numericz}, is sharply peaked in $(D-1)$ places; the peaks are between peaks of  $\mathcal{P}(V)$. Wiseman and Ralph have pointed out, for a qbit, that this arrangement leads to poorly purifying trajectories dominating $\expt{L(t)}$~\cite{WisRal06}. In a qbit the poorly purifying trajectories are those trajectories with eigenvalues of the same magnitude ($V(t)\approx 0$) in the long-time limit. Physically this means that the measurement has not projected, or the filter has not decided, if the state is in $z=\pm \half$. In qdits there are $(D-1)$ regions like this, where $V(t)\approx r$ for $r\in [J-\half,J-\smallfrac{3}{2},\ldots,J+\smallfrac{3}{2},-J+\half]$. In any of these regions only two eigenvalues contribute significantly to the integrand. Consequently it is reasonable to truncate the state to the two eigenvalues in a particular region and renormalize.

The effect of the two eigenvalue approximation is two-fold: it increases the purity of the state, and makes it possible to derive a fully analytic expression for the impurity. The eigenvalues of truncated state matrix are $(\lambda_0,\lambda_1)/(\lambda_0+\lambda_1)$\footnote{The order and placement of the eigenvalues on the diagonal will depend on the initial state.}, where $\lambda_0$ is the largest eigenvalue of $\rho$ and  $\lambda_a>\lambda_b$ when $a<b$. It is easy to show that $\rho$ is majorized \cite{NieChu00} by $\rho_2$ (that is, $\rho \prec \rho_2$), meaning that the original state is more mixed than our two eigenvalue approximation to it. From the fact that the purity is Schur-convex \cite{Bha96} it follows that $\tr{\rho^2}\le \tr{\rho^2_{\mathrm{2}}}$. This means the impurity of the two eigenvalue approximation is a lower bound on the true impurity: $L(\rho)\ge L(\rho_{\mathrm{2}} )$. In the long-time limit it is reasonable to expect that $L(\rho) \approx L(\rho_{\mathrm{2}} )$ from the arguments above.   

Now we calculate the impurity of the truncated state matrix $\expt{L_2(t)}$. We split the integral in \erf{impint} into regions such that the same two eigenvalues are the largest two eigenvalues for all values of $V$ in each region. The integration is then performed in each of these regions and then the regions are summed. Figure (\ref{regions}) illustrates how the regions should be split for $D=5$. After a coordinate transform (for example in region I it would be  $V' = V-1/2$) the integral in region one becomes
\begin{equation}\label{region1}
  \mathfrak{R}_{\mathrm{I}} = \frac{e^{-\gamma t}}{D}\sqrt{\frac{4\pi t}{\pi}} \int_{-\half}^{\half} \frac{e^{-4\gamma t V'^2}}{\cosh{(4\gamma t V')}} dV'.
\end{equation}
After a similar transformation the integral in region II becomes
\begin{equation}\label{region2}
  \mathfrak{R}_{\mathrm{II}} = \frac{e^{-\gamma t}}{D}\sqrt{\frac{4\pi t}{\pi}} \int_{-\half}^{\infty} \frac{e^{-4\gamma t V'^2}}{\cosh{(4\gamma t V')}} dV'.
\end{equation}
The total integral is thus
\begin{equation}\label{intregions}
\expt{L_2(t)} =  2\mathfrak{R}_{\mathrm{II}} + (D-3)\mathfrak{R}_{\mathrm{I}}.
\end{equation}
For $t\gg \gamma^{-1}$ the integrands become sharply peaked about $V'=0$ with negligible contributions to the integrals from the region outside the interval $V'\in[-\half,\half]$, so that $\mathfrak{R}_{\mathrm{II}}\simeq \mathfrak{R}_{\mathrm{I}}$. In this limit we have
\begin{equation}\label{nfb_2_fullint}
   \expt{L_{2}(t)} \simeq (D-1)\frac{e^{-\gamma t}}{D}\sqrt{\frac{4\pi t}{\pi}} \int_{-\half}^{\half} \frac{e^{-4\gamma t V'^2}}{\cosh{(4\gamma V')}} dV'.
\end{equation}
The final approximation is arrived at by noting that in the long-time limit the distribution in the numerator is broad compared to the distribution in the numerator for $t\gg \gamma^{-1}$. Thus we can make the approximation: $\int_{-\infty}^{\phantom{..}\infty} dVe^{-4\gamma t V'^{2}}/\cosh(\sqrt{2\gamma}V)\approx\int_{-\infty}^{\phantom{..}\infty} dV/\cosh(\sqrt{2\gamma}V)  =\pi/\sqrt{2\gamma}$. The final expression for the impurity is thus
\begin{equation}\label{nfb_2_full}
   \expt{L_{2}(t)}_{\mathrm{LT}} = \frac{2(D-1)}{D}\frac{\pi e^{-\gamma t}}{\sqrt{16\gamma t \pi}}.
\end{equation}
This is the analytic expression for $\expt{L(t)}$ that we set out to find, and is the key result of this section. 

Because previous work has used the exact qbit results we write it out for future reference:
\begin{equation}
  \expt{L(t)}_{\rm qbit} = \frac{e^{-\gamma t}}{2}\sqrt{\frac{4\pi t}{\pi}} \int_{-\infty}^{\,\infty} \frac{e^{-4\gamma t V'^2}}{\cosh{(4\gamma t V')}} dV'\label{nfb_qbit}.
  \end{equation}
  The long time limit of this expression is
  \begin{equation}
   \expt{L_{2}(t)}_{\mathrm{qbit -LT}} =\frac{\pi e^{-\gamma t}}{\sqrt{16\gamma t \pi}}\label{nfb_qbit_lt}.
\end{equation}
As could reasonably be expected for a qdit under the two-eigenvalue approximation (effectively a two-level system), the resulting average impurity is proportional to the qbit impurity with the proportionality depending on the qdit dimensionality. In Sec.~\ref{compareexact} we examine the accuracy and validity of this approximation.

\subsection{Comparison with numerics}\label{compareexact}
In \frf{cm2exact} (a) one can see that for $ t \gtrsim \gamma^{-1}$ the impurity, found by numerically evaluating \erf{impint}, is better approximated by \erf{intregions} than \erf{nfb_qbit}. Moreover, for all times the qbit impurity \erf{nfb_qbit} is a lower bound on the true impurity. However, when calculating the speed-up one is interested in times where $\gamma t\gg 1$, which are plotted in \frf{cm2exact} (b). 

\begin{figure}[h!]
\begin{center}
\leavevmode \includegraphics[width=\hsize]{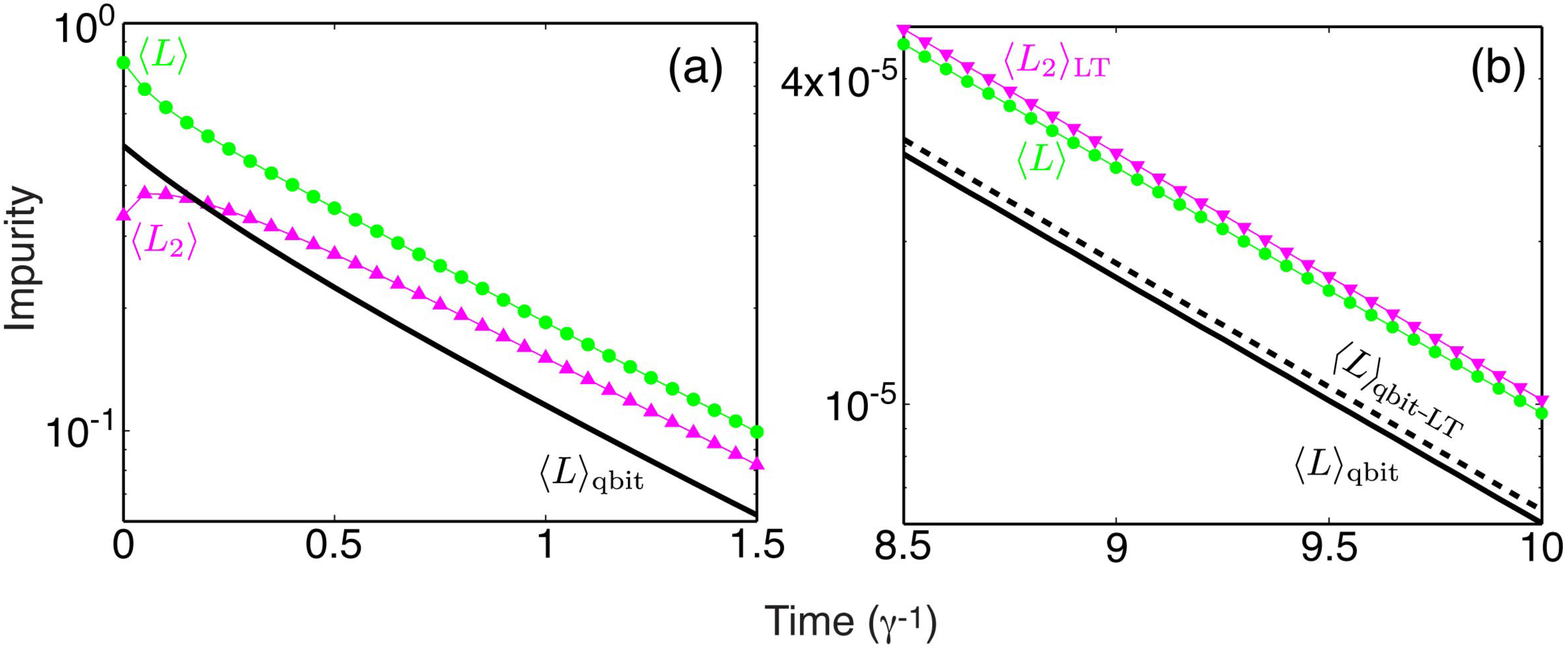}
\caption{A comparison of approximations to the evolution of $\expt{L}$, for $D=3$. (a) (starting from the upper-most curve): $\expt{L}$ \ie~\erf{impint} shown by green circles; $\expt{L_2}$ \ie~\erf{intregions} with the integrals evaluated numerically, shown by  magenta upward pointing triangles; $\expt{L}_{\mathrm{qbit}}$ \ie~\erf{nfb_qbit} shown by solid black line. (b) (starting from the upper-most curve): $\expt{L_2}_{\mathrm{LT}}$ \ie~\erf{nfb_2_full} shown by magenta downward pointing triangles; $\expt{L}$ \ie~\erf{impint} shown by green circles; $\expt{L}_{\mathrm{qbit-LT}}$ \ie~\erf{nfb_qbit_lt} shown by dashed black line; $\expt{L}_{\mathrm{qbit}}$ \ie~\erf{nfb_qbit} shown by solid black line. \erf{intregions} is not plotted in (b) because on this scale it is indistinguishable from $\expt{L}$.} \label{cm2exact}
\end{center}
\end{figure}

\erf{intregions} is not plotted in  \frf{cm2exact} (b) because on this scale it is practically indistinguishable from $\expt{L}$ at long-times (it approaches $\expt{L}$ from below). But $\expt{L_2(t)}_{\mathrm{LT}}$, that is \erf{nfb_2_full}, approaches $\expt{L}$ from above. Thus the approximations made to obtain \erf{nfb_2_full} from \erf{intregions} increase the impurity. This effect is easily understood once one realizes that the approximations amount to: throwing away parts of the integral \erf{nfb_2_fullint}; and over-estimating the integral by approximating $\exp{(-4\gamma t V'^2)=1}$. The next closest curve is that of the qbit long-time limit expression \erf{nfb_qbit_lt}, and below that is the qbit exact result \erf{nfb_qbit}. 

From these curves we infer the following: the long-time expression for the two eigenvalue approximation, $\expt{L_2(t)}_{\mathrm{LT}}$, is closer to $\expt{L}$ than the qbit long-time limit; $\expt{L_2(t)}_{\mathrm{LT}}$ is approaching $\expt{L}$ in the same way $\expt{L(t)}_{\mathrm{qbit-LT}}$ is approaching $\expt{L(t)}_{\mathrm{qbit}}$. This is important, as previously in the literature it has been common to use the qbit long-time limit expression as a lower bound for $\expt{L}$ any dimension \cite{ComJac06,ComWisJac08}. It is clear that for all times $t\gg\gamma^{-1}$,  $\expt{L_2(t)}_{\mathrm{LT}}$ is a better approximation to $\expt{L}$ than the qbit expression. Of course asymptotically the expressions decay at the same rate. Therefore there is nothing in this work to suggest that the speed-up calculated in previous works are incorrect.

\subsection{Stochastic simulations}\label{compare_stoc}
We now compare the analytic solutions from Section \ref{L_nfb_sec_analysis} and the exact numerics from Section \ref{compareexact}, to stochastic non-linear trajectory simulations of \erf{SME}. We used an Euler integration method with the following parameters: $\delta t=1\times 10^{-4}\gamma^{-1}$; $D=5$; ensemble size $=20$. The ensemble size is small because we will plot all of the trajectories so that we may gain some qualitative understanding from them.

\begin{figure}
\begin{center}
\leavevmode \includegraphics[width=\hsize]{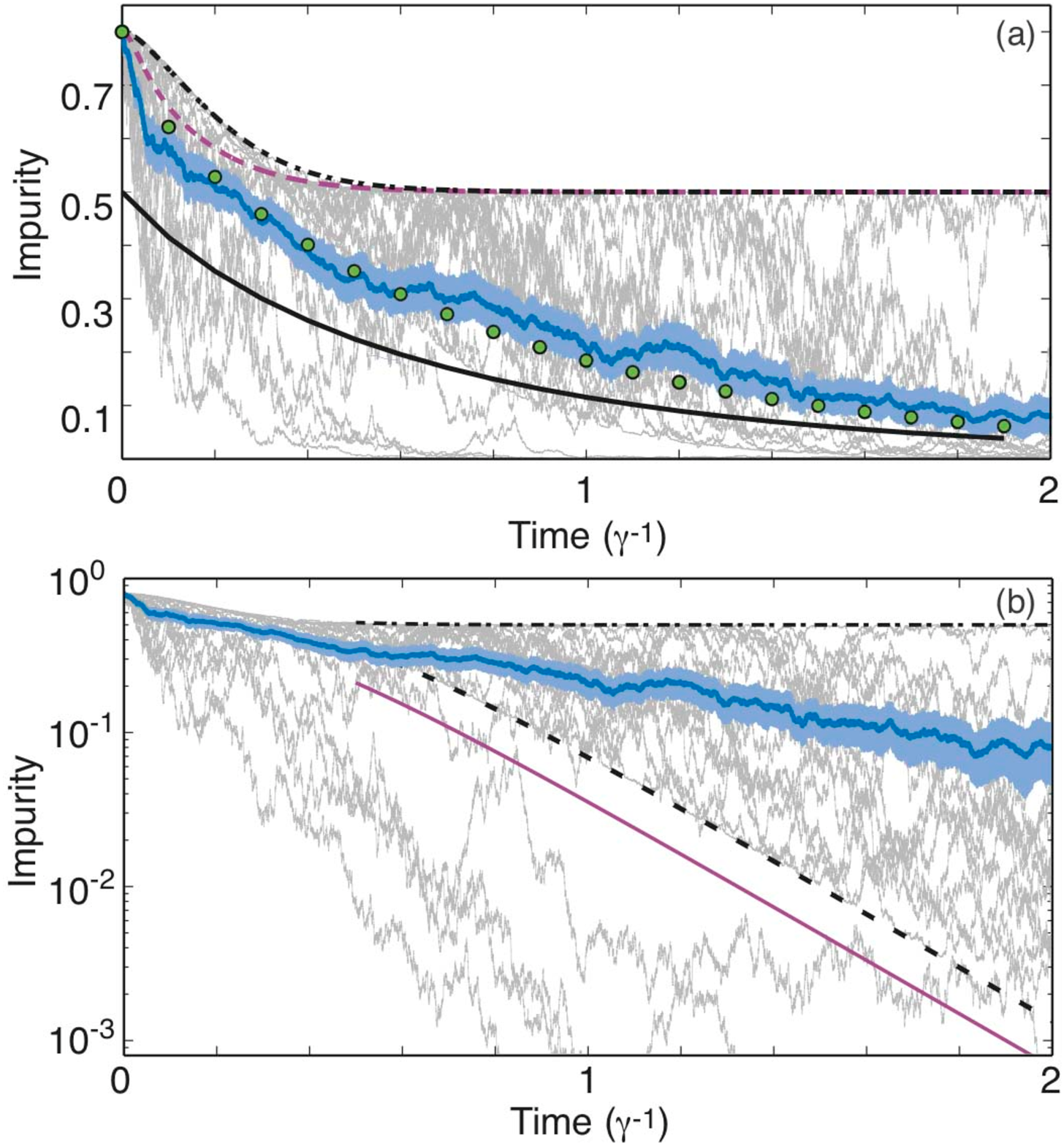}
\caption{Short time impurity dynamics for $D=5$. We plot $20$ impurity trajectories (grey); the  ensemble average of the trajectories (blue); the error bars for the ensemble average, $\pm  \sigma_{\mathrm{mean}}$, are smeared around the mean (light blue). Time is measured in units of $\gamma^{-1}$. Fig. (a) Linear abscissa. The dashed and dot-dashed lines are approximate upper bounds on the $L$ trajectories; see \srf{SEC_explain} for further explanation. The solid black line is the linear trajectories solution for the qbit, that is \erf{nfb_qbit}. The circles are $\expt{L}$ for $D=5$, \ie\ \erf{impint}.  Fig. (b) the impurity is plotted on a logarithmic axis; this helps to see the approximately exponential decrease in impurity predicted by \erf{nfb_2_full}. The approximate upper bound on the spread of $L$ trajectories is the dot dash line. The dashed line corresponds to the most likely trajectories. Below the solid (magenta) line there is only $1/D$ of the ensemble of trajectories at long-times. {These lines are explained in \srf{SEC_explain}.}} \label{cmtraj}
\end{center}
\end{figure}
 
Figure \ref{cmtraj} (a) shows that the $L$ of all of the trajectories is bounded from above by $\half$ for $t\gg \gamma ^{-1}$. Intuitively this effect can be understood from the two eigenvalue approximation; in the long-time limit a poorly purifying trajectory has two eigenvalues and thus these trajectories are upper bounded by $L=\half$. A more rigorous explanation is given in Sec.~\ref{SEC_explain}. On this Figure we also plot the numerically calculated ensemble average impurity and the linear trajectories solution, \erf{impint}.

There is some difficulty in obtaining convergence for $\expt{L}$ in the stochastic simulations, even with very large ensemble sizes. The convergence problem is due to a small number of poorly purifying trajectories first noted for a qbit in Ref.~\cite{WisRal06}. These poorly purifying trajectories are evident in \frf{cmtraj} (b) in that many trajectories touch the dot-dashed line at $L=\half$. For moderate ensemble sizes the whole region of possible impurities is filled by the ensemble even at $t =2\gamma^{-1}$. As noted above the major contributions to $\expt{L}$ are from the regions where $V = r $ (see \frf{regions}); that is, for results $V$ that are as far as possible from the most likely values of $V$ [the peaks of $\mathcal{P}(V)$]. Thankfully one can use \erf{impint} to calculate the evolution of the mean impurity. The advantages of this method are two fold: the computation time is greatly reduced, and it is exact, to numerical precision. 

Figure \ref{cmtraj} (b) shows that many trajectories also touch the dashed line corresponding to a scaling  $L \sim \exp{(-4\gamma t)}$. The reason for this clustering, and that at $L=\half$, will be explained in Sec.~\ref{SEC_explain}.


\subsection{The distribution of impurities}\label{SEC_explain}
In  Ref.~\cite{StoVanMab04}   Stockton, van Handel, and Mabuchi gave reasonable but non-rigorous arguments for a bound   [their Eq.~(41)]   on the degree of spreading of trajectories (that is, on the width of the distribution of trajectories) for the model of \erf{SME}. We now use linear trajectories to rigorously derive the bounds that they found. In all cases it is easy to obtain an analytic expression, but the exact form of the expression depends on $D$. Therefore we quote only the {\em long-time} asymptotic scaling of the bounds.

Central to our explanation of the bounds of the distribution of trajectories is the intuition gained by examining \frf{logregions}, where we have replotted portions of \frf{regions} on two separate axes. In \frf{logregions} (a) $\mathcal{P}(V,t)$ is plotted. This probability density plot is significant in the current analysis because it shows which records, $V$, are likely  (the peaks of $\mathcal{P}$) and which are unlikely (the troughs). In \frf{logregions} (b) the kernel $\Lambda(V,t)$, that is \erf{numericz}, is plotted on a log scale. The key features of this plot are: 1) the peaks at $V=-\smallfrac{3}{2},-\smallfrac{1}{2},\smallfrac{1}{2},\smallfrac{3}{2}$ which represent very impure trajectories and correspond to unlikely records, and 2) the troughs at $V=-1,0,1$ coincide with some of the purest trajectories and correspond to minima of $\mathcal{P}(V,t)$. We will elaborate on other relevant features of this graph below as each bound is explained.

\begin{figure}[h!]
\begin{center}
\leavevmode \includegraphics[width=\hsize]{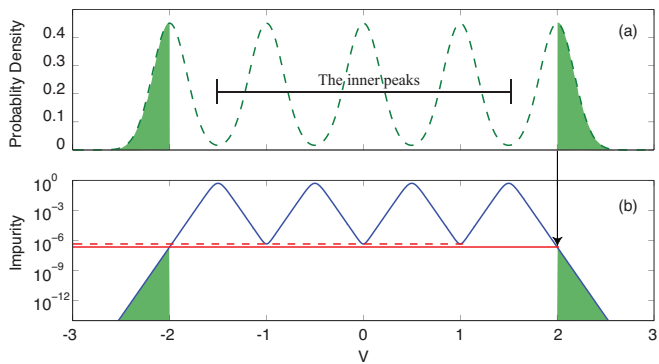}
\caption{(a) probability distribution $\mathcal{P}(V,t)$, \erf{probdisjz_scaled}, for $D=5$ and $t=4\gamma^{-1}$. The shaded regions of the probability distribution contain $1/D$ of the total probability at long-times. (b) the solid blue line is $\Lambda(V,t)$, \erf{numericz}, plotted with logarithmic scale on the abscissa for $D=5$ and $t=4\gamma^{-1}$. The red dashed line is a projection onto the abscissa of the minima of $\Lambda(V,t)$ at $V=-1,0,1$. The red solid line is a projection onto the abscissa of the $\Lambda(V,t)$ evaluated at $V=\pm 2$; notice that this projection is purer than the dashed line. The shaded regions in this figure depict the impurity of records that have $|V|\ge J$. As these regions concur with the regions in Fig. (a) they can be interposed as saying that only a proportion $1/D$ of the trajectories will have impurities smaller than those given by the projection $\Lambda(\pm J,t)$.} \label{logregions}
\end{center}
\end{figure}

1.{ \it The upper bound:} This is obtained by substituting $V(t)= r$ (recall that $r\in [J-\half,J-\smallfrac{3}{2},\ldots,J+\smallfrac{3}{2},-J+\half]$) into \erf{rhodiag}, normalizing, and then calculating the impurity. This procedure gives the upper bound because these records correspond to the most impure trajectories, that is peaks in \frf{logregions} (b). Because this procedure also corresponds to a a worst case scenario of the two eigenvalue approximation it can also be interpreted as the filter being unable to decided between the eigenvalues $\lambda_{r-\half}$ and $\lambda_{r+\half}$. For $D=5$ the records of interest correspond to $V=-\smallfrac{3}{2},-\smallfrac{1}{2},\smallfrac{1}{2},\smallfrac{3}{2}$ (see \frf{logregions} and \frf{regions}). In the long-time limit all solutions give 
\begin{equation}\label{L_LT_upperbound}
L_{\rm LT}=\half,
\end{equation}
which is also the bound predicted by Stockton and coworkers~\cite{StoVanMab04}. For small times there is some variation between the bounds given by $V=r=\pm\half$ and $V=r=\pm \smallfrac{3}{2}$. This is because $\mathcal{P}(V)$ is broad at short-times so that more eigenvalues contribute to the impurity when for example $V=\half$. Conversely when $V=\smallfrac{3}{2}$ the is only one eigenvalue to the left of  this record. Accordingly it is reasonable to expect that at short-times the bound given by $V=\pm\half$ will produce a bound that is more impure than the bound given by $V=\pm \smallfrac{3}{2}$. This behaviour is confirmed in \frf{cmtraj} (a), where the dashed line is $V = \pm \smallfrac{3}{2}$ and the dot dashed line is $V=\pm\half$. 

2.{ \it A pseudo lower bound:}  When $V(t)= s$ for $s\in [J-1,\ldots,-J+1]$, the record corresponds to the inner peaks of the probability distribution, as depicted in \frf{logregions} (a) (the peaks at $\pm J$ are {\em not} included in this analysis). These peaks of $\mathcal{P}(V)$ are the most likely records and coincide with some of the purest trajectories, the minima in \frf{logregions} (b). For $D=5$ the peaks of interest are those corresponding to $V=-1,0,1$ in \frf{logregions}. By substituting $V(t) = s$ into \erf{rhodiag}, normalizing, and then calculating the impurity, one obtains the lower bound without any approximations. These solutions can be obtained for any dimension $D$. By making a two-eigenvalue approximation one can solve the evolution of the impurity analytically giving 
\begin{equation}\label{L_LT_lowerbound}
L_{\rm LT} \sim \exp{(-4\gamma t)}.
\end{equation}
Equation~(\ref{L_LT_lowerbound}) corresponds to the lower bound quoted by Stockton and coworkers  \cite{StoVanMab04}. In their case, their bound appeared to be a true lower bound as their initial state was a coherent spin state (a collection of $D/2$ spin one half particles where $D=2J+1$ with mean spin vector, $J$, aligned along the $x$-axis in this case). This has a small population in $\ket{\pm J}$, and it is therefore unlikely in their case that $|V|>|J|$. For our analysis we see clearly that the bound is not a true lower bound (see \frf{cmtraj} (a)), because a maximally mixed state has equal populations in all eigenstates. However, because these records $V(t)=j$ are the most likely records many of the trajectories touch or cluster around this bound. This will be further explored in point 4.~below.

3.{ \it Physically likely trajectories:\label{phys_like}} In our case there is no lower bound, although a trajectory that purifies infinitely fast is infinitely unlikely~\footnote{ P. E. F. Mendon\c{c}a and J. Combes, unpublished, 2006.}. It is possible to give a natural bound on the physically likely trajectories based on the probability distribution $\mathcal{P}(V)$. In the lower bound section above we did not include the peaks of the probability distribution found at $V=\pm J$. It is clear from \frf{logregions} (b) that there is no corresponding minimum of the kernel $\Lambda(V,t)$ in this region. When $\Lambda(V,t)$ is evaluated at $V(t)=\pm J$ one finds that it gives impurities smaller (see the solid red line in \frf{logregions} (b)) than those which correspond to the inner peaks, [the dashed red line in \frf{logregions} (b)]. From the shaded regions in \frf{logregions} (a) it is apparent that only $1/D$ of the probability density is found for $|V|\ge J$. This means that, in the long-time limit, only a proportion $1/D$ of the trajectories will have purities greater than the bound we now present. Using the linear trajectory solution with $V(t)=\pm J$ and solving for the impurity one obtains the bound. By making a two eigenvalue approximation one finds that the bound scales asymptotically as 
\begin{equation}\label{L_LT_physlikely}
L_{\rm LT} \sim \exp{(-4\gamma t)}.
\end{equation}
The scaling in \erf{L_LT_physlikely} and \erf{L_LT_lowerbound} is precisely the same. The difference in the bounds is only apparent when the calculation is exact; see \frf{cmtraj} (b). The dashed line (black) is \erf{L_LT_lowerbound}, while the thin solid line (magenta) is \erf{L_LT_physlikely} (also see \frf{cmtrajexp}).   Obviously for a qbit ($D=2$) the bound given by $V=\pm \half $ corresponds to the median of the distribution of impurites, denoted by $\wp(L,t) dL$, at all times, and the mode at long-times. 

4.{ \it Distribution of Impurities:} 
The distribution of impurites can be calculated analytically for $D=2$ \cite{WisRal06}. However we will change variables to the log-impurity so that we may clearly see the featured discussed above. The required change of variables from Eq.~(24) in Ref.~\cite{WisRal06} is $\ell\equiv\log_{10} L = \log_{10} {\left [\half (1-z^2)\right ]}$ so that $z=\sqrt{1-2\times10^{\ell}}$. The distribution of log-impurites is thus
\begin{eqnarray}\label{dis_imps_qbit}
d\ell \wp(\ell,t)\!\!&=& \!\!\nn d\ell\frac{e^{-\gamma t}\ln{(10)}\cosh{\!\left [\mathrm{arctanh}^2(\sqrt{1-2\times10^{\ell}})\right ]}} {2\sqrt{8\pi\gamma t}\sqrt{1-2\times10^{\ell}}} \\
&&\, \times  \exp{\left(-\mathrm{arctanh}^2(\sqrt{1-2\times10^{\ell}})/4\gamma t\right)}.
\end{eqnarray}
In \frf{cmtrajexp} this distribution is plotted at $t=2\gamma^{-1}$ (the red dashed curve). The peak of this distribution at ${\ell} =-0.301$ becomes less prominent at long-times. This is when, as Wiseman and Ralph pointed out \cite{WisRal06}, considering the average log-impurity is a good way to find an approximation to the mean time to a fixed $L$.

The poorly purifying trajectories affect $\expt{L}$ and a large number of these trajectories cluster around the upper bound (as seen in \frf{cmtraj}). This motivates the consideration of another measure of mixedness which de-emphasises the peaks of the distribution \cite{WisRal06,ComWisJac08}. However, in this article, we will remain focused on $\expt{L}$.

\begin{figure}[h!]
\begin{center}
\leavevmode \includegraphics[width=\hsize]{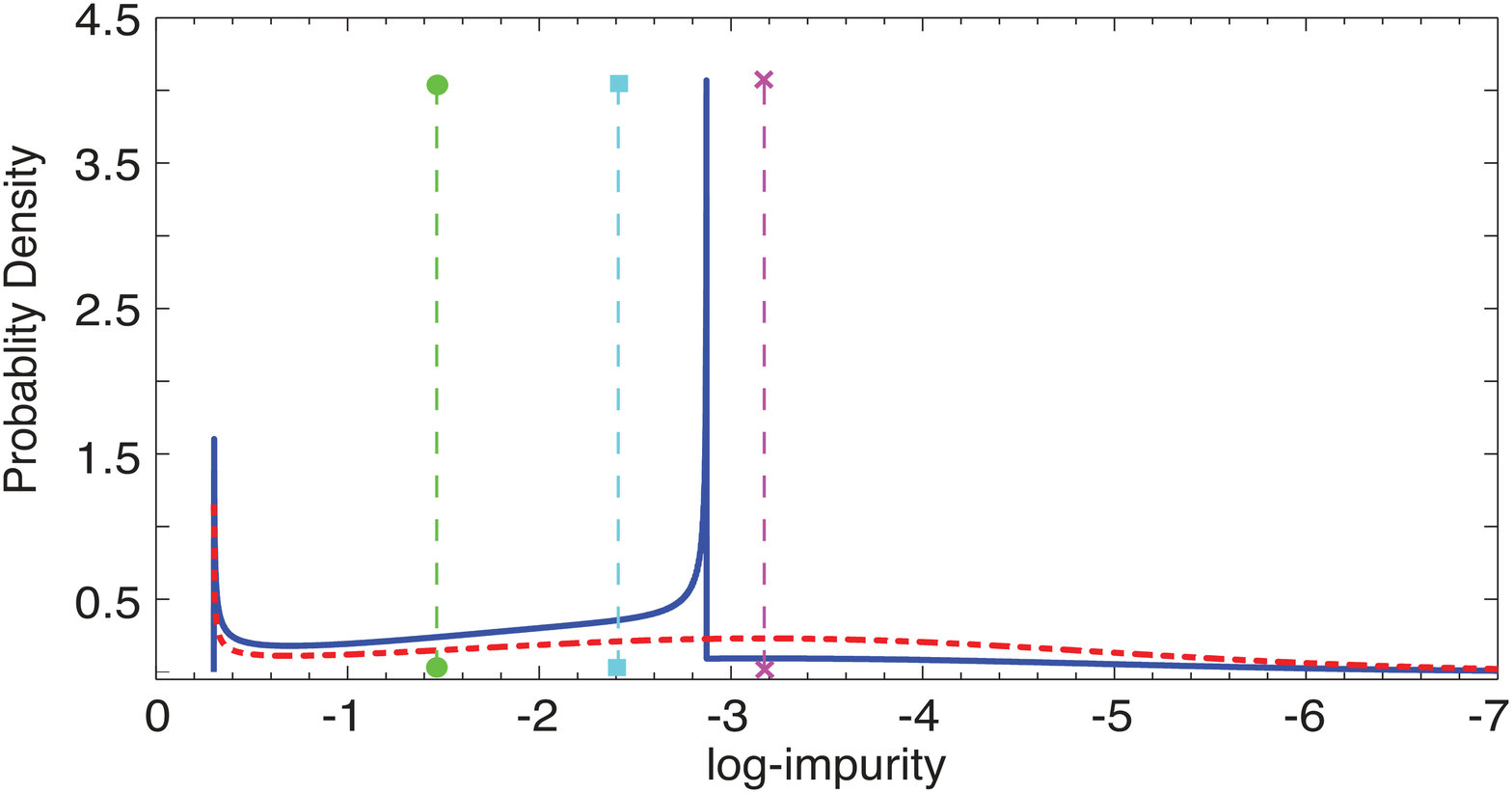}
\caption{The distribution of log-impurities at $t=2\gamma^{-1}$. The blue solid curve is the distribution for $D=5$. The red dashed curve is for $D=2$, that is \erf{dis_imps_qbit}. The line with the circles is $\log_{10}(\expt{L}) =-1.46$ \blk at $t=2\gamma^{-1}$, according to \erf{impint}; the line with the squares is $\expt{{\ell}}= -2.41$ at $t=2\gamma^{-1}$. Only a proportion $1/D$ of the trajectories have lower purity than (to the right of) the line with the crosses (plotted at $\ell= -3.1736$) as explained in the main text. The two other features on the graph at ${\ell}=   -0.3010$ and $\ell=-2.87$ correspond to the upper and the pseudo-lower bounds on the distribution of impurities described in the main text. } \label{cmtrajexp}
\end{center}
\end{figure}

For $D>2$ the distribution of ${\ell}$ can be numerically calculated from the probability distribution $\mathcal{P}(V)$ and $\Lambda(V,t)$ as follows. First one splits the distribution $\Lambda(V,t)$ into regions $\{(-\infty,-j),[-j,-j+\half),\ldots,[j,\infty) \}$. These regions are chosen because ${\ell}$ is always monotonically increasing or decreasing in these regions; see \frf{logregions} (b). This is required to perform the numerical interpolation required by the following procedure. In each region one discretizes the range of ${\ell}$ and then finds the values of $V$ that correspond to this ${\ell}$. The probability distribution $\mathcal{P}(V)$ is then integrated over this region  to give $d\ell \wp(\ell,t)$. This is then summed over all regions. 

In \frf{cmtrajexp} the solid curve is the probability distribution for $D=5$ at $t=2\gamma^{-1}$; this distribution was obtained using the above method. The upper bound and pseudo-lower bound described in the text above are quite apparent in distribution; they correspond to the two sharp features at ${\ell}=-0.301$ and ${\ell}=-2.87$. It is also apparent that $\expt{{\ell}}$ is a more faithful central tendency measure than $\log_{10}\expt{L}$  as the line corresponding to $\expt{{\ell}}$ is closer to the bulk of the distribution.

\subsection{Summary}\label{nfbsum}
We conclude our study of information extraction from commuting measurements by summarising the key points. It is possible to characterise the average amount of information extracted using the average impurity. We found a simple expression for the average impurity, $\expt{L_{2}(t)}_{\rm LT}  \sim \exp(-\gamma t)$,  which is valid when $t\gg \gamma ^{-1}$. Additionally we found that in this asymptotic limit a portion $(1-1/D)$ of the trajectories would be bounded by $c \exp(-4\gamma t) \le L \le \half$ for a constant $c$ that we can determine for a particular $D$. It is the existence of rare but poorly purifying trajectories which 
explains the upper bound of $\half$, and which explains the difference between the scaling of the mean 
and that of the lower bound, $c \exp(-4\gamma t)$. Indeed, other measures of central tendency, such as the median, or the exponential of the mean-log,  exhibit the same scaling as the lower bound, $\sim \exp(-4\gamma t)$,  which is also reflected in the mean time to attain a given purity \cite{ComWisSco10}.

Despite its limitations, we have used $\expt{L}$, which scales as 
$\exp(-\gamma t)$, as our measure of average information in the main text of the paper. Here we briefly comment 
on what would change if we were to have used a more faithful measure of central tendency such as the median. \blk 
In \srf{nopermsec} we found that the spread of trajectories for the QFT protocol was small for sufficiently frequent feedback. This feature is true in general for measurement in an unbiased basis \cite{ComWisJacOCo10}. The main consequence of this, for complementary continuous measurements, is the mean impurity is approximately equal to the median impurity. 
Thus we can directly compare the results obtained for $\expt{L}$ in the main part of the paper to the scaling 
for the median impurity, $\sim \exp(-4\gamma t)$, for commuting measurement. The result, obviously, 
is a diminution in the speed-up offered by the latter by a factor of four: $\smallfrac{1}{6}(D+1)  \le S^{\rm median} \le  \smallfrac{1}{8} D^{2} $. For small systems ($D=2$ or 3) this implies a slow-down, or at best a modest speed-up. But for large $D$ the scaling with $D$ implies that using feedback to construct suitable measurements in an unbiassed basis will beat the no-feedback protocol even in terms of the median impurity or mean time.

\section{Information acquisition in dimension $D$ and its relationship to complementarity}\label{qditqft}


\subsection{Previous results}\label{L_Dfb_sec}
In this subsection we briefly summarise the results of previous studies on information acquisition by measurement in a complementary basis. The original articles on the subject are found in Refs.~\cite{FucJac01,Jac0303,ComJac06,ComWisJacOCo10}. The basic idea is to use quantum feedback control to continuously keep the eigenbasis of the state complementary to the measurement basis. By analogy with the intuition presented in \frf{fig0} it is hoped that this procedure will enhance information extraction.

It is easy to show, from \erf{SME}, using It\={o} calculus that the information acquired~\cite{FucJac01}, as characterised by the change in impurity, about the state due to measurement of infinitesimal duration is~\cite{ComWisJacOCo10}
\begin{eqnarray}\label{dlnfb}
  \nonumber dL&=&- 8\gamma dt \{ \tr{\rho X\rho X}-2\tr{X\rho}\tr{X\rho^2}\\&&\nn\phantom{- 8\gamma dt} \,+\tr{\rho X}^2\tr{\rho^2}\}\\
 &&-4\sqrt{2\gamma}dw\{\tr{X\rho^2}-\tr{\rho X}\tr{\rho^2}\}.
\end{eqnarray}
We wish to study the effect of measuring in a complementary basis. In order to measure in a complementary basis through out the measurement process one must use quantum feedback control \cite{WisMil10} to continually adjust the basis. We will not labour on those details here as they have been adequately discussed before \cite{Jac0303,ComJac06,ComWisJacOCo10,WisMil10}. The required transformation of the measurement basis is $X(t)\rightarrow \check{X}(t)$ such that $|\ip{\check{x}}{i}|=1/\sqrt{D}$, for all $\check{x}$ and $i$ where $\ket{i}$ is an eigenstate of $\rho$ and $\ket{\check{x}}$ is an eigenstate of $\check{X}$. Replacing all the $X$'s by $\check{X}$ in \erf{dlnfb} considerably simplifies the expression to \cite{ComJac06, ComWisJacOCo10}
\begin{eqnarray}\label{dgen_dpurity}
 dL&=&- 8\gamma\tr{\check{X}\rho \check{X}\rho}dt,
\end{eqnarray}
or equivalently 
\begin{eqnarray}\label{gen_dpurity2}
 dL&=&- 8\gamma dt \sum_{r,c} |\check{X}_{r,c}|^2\lambda_r\lambda_c.
\end{eqnarray}
Observe that the unbiased condition which simplified \erf{dlnfb} to \erf{dgen_dpurity} leaves a permutational degree of freedom in Eqs.~(\ref{dgen_dpurity}) and~(\ref{gen_dpurity2}) because a permutation of the eigenstates of $\rho$ does not affect $|\ip{\check{x}}{i}|=1/\sqrt{D}$. This means we should optimise over permutations in the expression $ dL=- 8\gamma\tr{\check{X}P_{m}\dg\rho P_{m} \check{X}P_{m}\dg\rho P_{m}}dt$, where $P_{m}$ is a permutation matrix, to maximize the decrease in impurity. To obtain a lower bound $|dL|$ of the optimal permutation the we average over all possible permutations: $ dL=- 8\gamma\sum_{m}\tr{\check{X}P_{m}\dg\rho P_{m} \check{X}P_{m}\dg\rho P_{m}}dt$. The average gives a lower bound because the sum of a sequence is always less than or equal to the greatest term in the sum.

It was shown in Refs.~\cite{ComJac06,ComWisJacOCo10} that a tight lower bound on \erf{dgen_dpurity}, for the optimal permutation, is
\begin{eqnarray}\label{dl_lbounds}
dL\le- \frac{2}{3}\gamma t (D+1) L(t).
\end{eqnarray}
While the upper bound was recently found to be
\begin{eqnarray}\label{dl_ubounds}
dL\ge -\frac{1}{2}\gamma t D^{2} L(t),
\end{eqnarray}
for $D\gg1$ \cite{ComWisJacOCo10}. Equations (\ref{dl_lbounds}) and (\ref{dl_ubounds}) hold for measurement in {\em any} complementary basis. Using these bounds one may compare the time it takes for a commuting measurement (\ie\ \erf{nfb_2_full}) to extract a certain amount of information (i.e. to attain $\expt{L} = \epsilon$) to how long a complementary measurement may take to extract that same amount of information. That is we equate \erf{nfb_2_full} and the solutions of Eqs. (\ref{dl_lbounds}) and (\ref{dl_ubounds}), 
and solve for the ratio of $t_{\rm complementary}$ and $t_{\rm commute}$. We call this ratio the speed-up in information acquisition. For $t_{\rm commute}\gg \gamma^{-1}$ the speed-up is bound by
\begin{eqnarray}\label{speedup_bounds}
\frac{2}{3}(D+1)\le S \le \frac{D^2}{2},
\end{eqnarray}
for $D\gg 1$ \cite{ComWisJacOCo10}.

While these results show the usefulness of feedback in rapid purification, there are some important issues that have not yet been addressed. First, does an explicit construction of such a complementary measurement strategy exist? Second, is actual performance of such a strategy 
$O(D)$ or $O(D^2)$? Third, what is the mechanism of the speed-up? The first and second questions are answered in \srf{nopermsec} and \srf{SEC_QFT_bounds} respectively. We defer the third question until \srf{whycomp}.

\subsection{Complementarity via quantum Fourier transform for $D=2,3,4$}\label{nopermsec}
In this subsection we explore implementing a continuous complementary measurement in systems of dimension three and four. There are many unbiased bases, so to make our analysis concrete we must choose one. It has been noted elsewhere \cite{JacLun07} that for $D>3$ not all unbiased bases are equally good at reducing the impurity. Nevertheless we choose a particular complementary, the one generated by the quantum Fourier transform (QFT) of the logical basis. The construction of the continuous complementary measurement then consists of: calculationally diagonalizing the state matrix; calculationally ordering (permuting) those eigenvalues to maximize $|\expt{dL}|$ after a measurement; and applying the appropriate permutation and the QFT unitary to the state. This whole procedure is an example of quantum feedback control \cite{WisMil10}. The $D$-dimensional QFT can be represented by the matrix
\begin{eqnarray}
 T =\frac{1}{\sqrt{D}}\left( \begin{array}{c c c c c c}
1      & 1        & 1       &\ldots& 1\\
1      & q        & q^{2} & \ldots& q^{1(D-1)} \\
1      & q^{2} & q^{4}  &\ldots & q^{2(D-1)}\\
\vdots &\vdots & \vdots &\ddots\\
1      & q^{(D-1)}&q^{2(D-1)}&&q^{(D-1)^2}\\
\end{array} \right) ,
\end{eqnarray}
where $q = \exp{\left(2\pi i /D\right)}$. The matrix elements of $T$ are 
\begin{equation}
T_{rc}=\frac{1}{\sqrt{D}}\exp{\left(\frac{2\pi i}{D}(r\times c)\right )},
\end{equation}
where $r,c\in[0,\ldots,D-1]$.

For $D=2$ Jacobs' protocol~\cite{Jac0303} turns out to be equivalent to the QFT protocol. The transformation unitaries in both cases are
\begin{align}
T_{\rm Jac} &=\exp\left(i \smallfrac{\pi}{2}J_{y}\right)=\smallfrac{1}{\sqrt{2}}\left(\begin{array}{cc}\phantom{-}1 & 1 \\-1 & 1\end{array}\right)\\
T_{\rm QFT}&=\smallfrac{1}{\sqrt{2}}\left(\begin{array}{cc}1 & \phantom{-}1 \\1 & -1\end{array}\right)
\end{align}
The effect of these transformations on $dL$ is identical: $\check{X}=T_{\rm Jac}^{\phantom{\dag}} J_{z}T_{\rm Jac}^\dag=T_{\rm QFT}^{\phantom{\dag}} J_{z}T_{\rm QFT}^\dag=J_{x}$. Because this scheme is equivalent to Jacobs we will not analyse this $D=2$ case any further.

 For a system of dimension $D=3$, the transformed measurement is
\begin{eqnarray}
\check{X}= TJ_zT^\dag =\frac{1}{3}\left( \begin{array}{c c c}
0     & 1-q & 1-q^{*}\\
1-q^{*}      & 0& 1-q\\
1-q      & 1-q^{*}& 0\\
\end{array} \right),
\end{eqnarray}
where $q = \exp(2\pi i/3)$. 
Taking the state to be $\rho=\mathrm{diag}(\lambda_0,\lambda_1,\lambda_2)$, where $\lambda_r>\lambda_c$ when $r<c$ and solving the equation for $dL$, using the fact that $L = \sum_{r\neq c,c\neq r}\lambda_r\lambda_c$, we find
\begin{eqnarray}\label{dld3}
\nn dL&=&- 8\gamma dt \sum_{r,c} |\check{X}_{r,c}|^2\lambda_r\lambda_c\\
&=& -8\gamma dt \left(\frac{2}{3}(\lambda_0\lambda_1+\lambda_0\lambda_2+\lambda_1\lambda_2) \right)\label{d3weights} \\
&=& -8\gamma dt \frac{1}{3}L(t),
\end{eqnarray}
which coincides with the bound of \erf{dl_lbounds}. The above expression can be easily integrated to give
\begin{figure}
\begin{center}
\leavevmode \includegraphics[width=0.3\hsize]{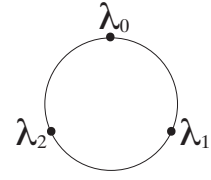}
\caption{The only possible permutation for $D=3$. The solid lines between the eigenvalues represent the weighting of the multiplying factors in Eq.~\ref{d3weights}. That is, $\smallfrac{2}{3} =|\check{X}_{r,c}|^{2}$, $\forall r,c$.}\label{d3perm}
\end{center}
\end{figure}
\begin{eqnarray}\label{ld3noperms}
 L(t)&=& e^{-\frac{8}{3}\gamma t }L(0),
\end{eqnarray}
In Fig \ref{d3perm} the weights $|\check{X}_{r,c}|^2$ between the $\lambda_r\lambda_c$'s are plotted. The weights in this case are all equal, so it is obvious that permuting the eigenvalues will not change the decrease of impurity. This explains why we did not have to find an optimal permutation.

Now we compare the above calculation to numerics. The average impurity for the continuous complementary measurement protocol and continuous commuting measurement are calculated by stochastic simulations and \erf{impint} respectively. Figure \ref{ld3fb} shows that the stochastic simulations of the complementary measurement protocol give a deterministically decreasing $\expt{L}$ as predicted by \erf{ld3noperms}. 

We numerically investigated the effect of varying the frequency, $1/\delta t$, of the operation applied (feedback) to keep the state and measurement eigenbases complementary in \frf{d3speed}. The stochastic fluctuations in the simulations arise from the finite size of $\delta t$ in the simulations. This effect can also be found in simulations of the Jacobs' qbit feedback protocol~\footnote{Combes and Jacobs August 2008, unpublished.}.  As $\delta t \rightarrow 0$ the numerically calculated average impurity for complementary measurement approaches \erf{ld3noperms}. For $\delta t \ll 10^{-3}\gamma^{-1}$ numerical simulations are indistinguishable from the analytically calculated (then interpolated) speed-up, the solid line in \frf{d3speed}. 
\begin{figure}
\begin{center}
\leavevmode \includegraphics[width=\hsize]{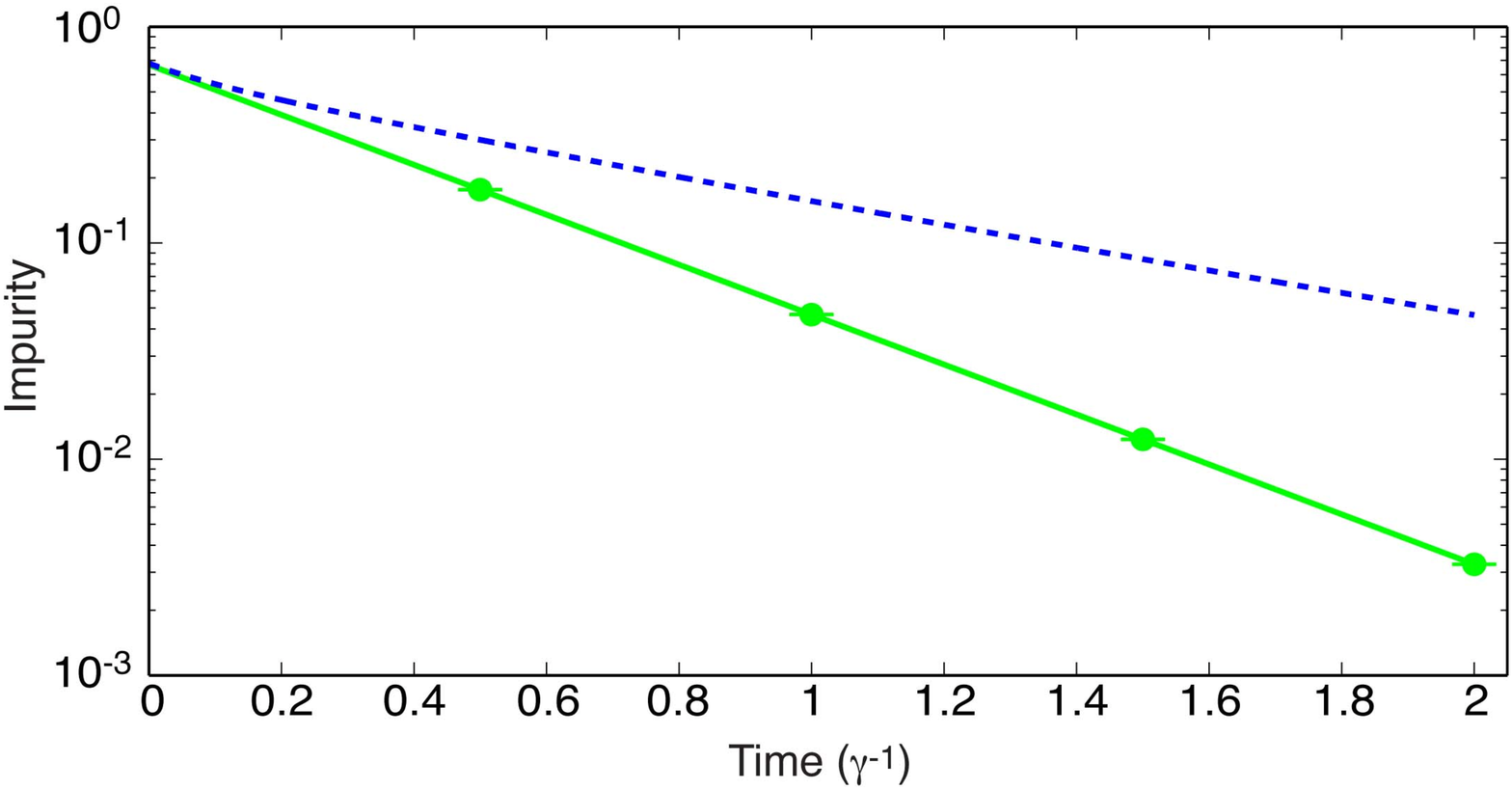}
\caption{Trajectory simulations of the impurity for a $D=3$ system subject to feedback every $\delta t = 10^{-3}$ $\gamma^{-1}$ to keep the state and measurement eigenbases complementary. The ensemble size is $100$. The dashed line is $\expt{L}$ for a continuous commuting measurement $(D=3)$ \ie\ \erf{impint}. The solid line is the numerically calculated ensemble average of $\expt{L}$ for the complementary measurement; it is plotted in green; error bars are plotted at times $(0.5,1,1.5,2)\times\gamma^{-1}$. The analytic expression for the average impurity, \erf{ld3noperms}, is indistinguishable from the results of the simulation.}\label{ld3fb}
\end{center}
\end{figure}

\begin{figure}[h!]
\begin{center}
\leavevmode \includegraphics[width=\hsize]{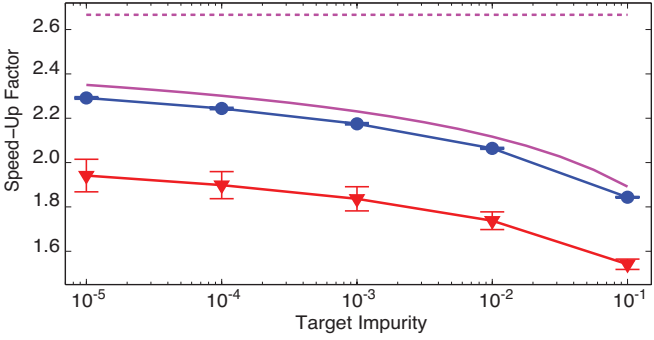}
\caption{Short time numerically calculated speed-up for $D=3$. The dashed line is the analytically calculated asymptotic speed-up $S= 8/3=2.\dot{6}$. The solid (magenta) line is the ideal speed-up; calculated numerically from equations  \erf{impint} and \erf{ld3noperms}. The circles and triangles are numerical calculations of the speed-up with finite $\delta t = 10^{-3}\gamma^{-1}$ and $10^{-2}\gamma^{-1}$ respectively; the ensemble size is $100$. As $\delta t \rightarrow 0$ the numerically calculated speed-up approaches the solid line and the error-bars vanish.} \label{d3speed}
\end{center}
\end{figure}

We now consider complementary measurement using the QFT for the case when $D=4$. The increment for the impurity is
\begin{eqnarray}\label{d4weights}
\nn dL&=&- 8\gamma dt \sum_{r,c} |\check{X}_{r,c}|^2\lambda_r\lambda_c\\
 &=&\nn -8\gamma dt \left(\lambda_0\lambda_1+\half \lambda_0\lambda_2+ \half \lambda_1\lambda_3 \right.\\&&\phantom{-8\gamma dt( }\left.+ \lambda_0\lambda_3 + \lambda_2\lambda_3+ \lambda_1\lambda_2 \right ).
\end{eqnarray}
\begin{figure}[h!]
\begin{center}
\leavevmode \includegraphics[width=\hsize]{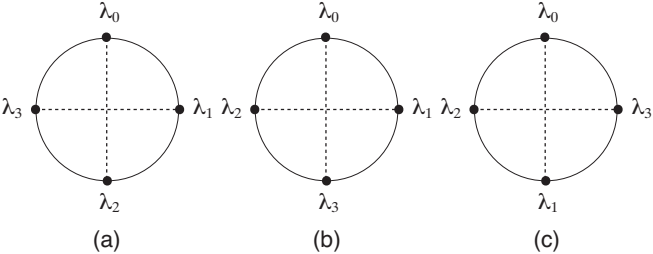}
\caption{(a) The weights $|X_{r,c}|^2$ in \erf{d4weights} after diagonalization and descending ordering. (b) The optimal permutation for $D=4$. (c) The worst permutation for $D=4$. The solid line represents maximum weighting of multiplication in \erf{d4weights}, the dashed lines represent minimum weights corresponding to $\half$. 
} \label{d4perm}
\end{center}
\end{figure}
Unfortunately, it is not possible to factor out an expression for the impurity on the RHS of the above equation. It is obvious from \erf{d4weights} that permuting the eigenvalues will affect $|dL|$. Our task is to maximize the decrease in $dL$. From \erf{d4weights} it is possible to intuit the form of the best and worst permutations, which are depicted in Fig. \ref{d4perm}. An optimal permutation is $\rho_{\mathrm{opt}}=\mathrm{diag}(\lambda_0,\lambda_1,\lambda_3,\lambda_2)$. This permutation is not uniquely optimal; the permutation $\mathrm{diag}(\lambda_1,\lambda_0,\lambda_2,\lambda_3) $ is also optimal. 
One of the worst permutations is $\rho_{\mathrm{worst}}=\mathrm{diag}(\lambda_0,\lambda_3,\lambda_1,\lambda_2)$. Unfortunately knowing the optimal permutation does not help us simplify the expression. Because we cannot solve this case analytically, we invoke the procedure developed in Refs.~\cite{ComWisJac08, ComWisJacOCo10} to find the bounds on $|\expt{dL}|$ (and hence $S$) for the QFT protocol for all $D$. These bounds are calculated in Section \ref{SEC_QFT_bounds}.

Figure~\ref{d4speed} depicts the numerically calculated speed-up for $D=4$. The speed-up is larger than that of the $D=3$ case \footnote{In the simulations we used a linear trajectory type simulation so that positivity of the state is ensured.}. Furthermore, for reasonably frequent feedback the numerically calculated speed-up lies between the analytical bounds predicted by \erf{dl_lbounds} and  \erf{dl_ubounds}.

\begin{figure}[h!]
\begin{center}
\leavevmode \includegraphics[width=\hsize]{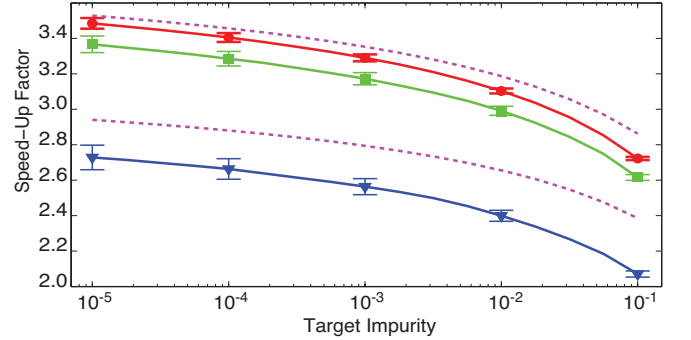}
\caption{Short time numerically calculated speed-up for $D=4$.  The dashed lines are the bounds on the speed-up given by \erf{dLf_qft} and \erf{eigenvalueising2}. The circles, squares and triangles are numerical calculations of the speed-up with finite $\delta t = 10^{-3}\gamma^{-1}, 10^{-2}\gamma^{-1}, 10^{-1}\gamma^{-1}$ respectively. To keep the error bars roughly the same size the ensemble size was varied. The ensemble sizes for circles, squares and triangles are $100$; $400$; and $6,400$ respectively.} .\label{d4speed}
\end{center}
\end{figure}
We also briefly compare the complementary measurement protocol to the rapid measurement (RM) algorithm of Ref.~\cite{ComWisJac08} in \frf{d4spread}. The decrease in the ensemble average of the impurity seems to be of the same order. However, the trajectories in the RM case have a large variance. {This is because the trajectories are not differentiable.} By comparison, even for a finite $\delta t$ the $L$ in complementary measurement is very close to {\em deterministic}. Although $L$ is still stochastic, it is also differentiable for $\delta t \rightarrow dt$ so the noise in the simulations is reduced. 
\begin{figure}[h!]
\begin{center}
\leavevmode \includegraphics[width=\hsize]{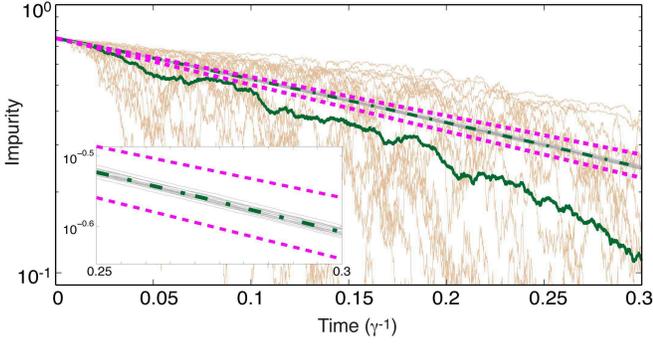}
\caption{A comparison between the complementary measurement protocol and rapid measurement (RM) feedback (of Ref.~\cite{ComWisJac08}) for $D =4$; $\delta t = 10^{-4}\gamma^{-1}$; and an ensemble of 20 trajectories in each case. The dashed lines are the analytically calculated bounds on $dL$ for complementary feedback [\erf{dLf_qft} and \erf{dL2_qft}]. The dot-dashed line is the ensemble average for $L$ with QFT feedback. The solid (dark green) line is the ensemble average for the RM feedback. The light brown lines are the trajectories for the RM feedback. The analytical bounds suggest that both feedback strategies have the same order of speed-up. Inset figure: all of the complementary measurement trajectories (grey lines) lie within the predicted bounds. The trajectories are very nearly deterministic.}\label{d4spread}
\end{center}
\end{figure}

\subsection{Bounds on the QFT complementary measurement protocol for all $D$}\label{SEC_QFT_bounds}
 
\subsubsection{Lower bound}\label{SEC_flat_QFT}

From \erf{speedup_bounds} it is clear that the lower bound on the {optimal} speed-up for measurement any complementary basis is $S= \smallfrac{2}{3} (D+1)$. Nevertheless it is interesting to work this out explicitly for a particular complementary basis. The method we use to find the lower bound was first presented in Ref.~\cite{ComWisJacOCo10}. This method uses a fictitious state $\rho_{F}$ with the same purity as the state under consideration, \ie\ $L[\rho]=L[\rho_{F}]$. The fictitious state $\rho_{F}$ is defined as $\rho_{ F} = {\rm diag} (1-\Delta, \delta,..,\delta)$ where $1-\Delta$ is the largest eigenvalue of $\rho_{F}$ and $\delta=\Delta/(D-1)$. The equation for $dL$ is invariant under permutations of the eigenvalues $\rho_{\rm F}$. Substituting $\rho_{\rm F}$ into the RHS of \erf{gen_dpurity2} gives
\begin{eqnarray}
\nn && 2(1-\Delta)\frac{\Delta}{(D-1)}\sum_{r\neq 1}|\check{X}_{r1}|^2+\frac{\Delta^2}{(D-1)^2}\sum_{r\neq c,c\neq r}|\check{X}_{rc}|^2.
\end{eqnarray}
This expression is further simplified by explicitly calculating the matrix elements. For the QFT we have
\begin{equation}
\sum_{r\neq 1}|\check{X}_{r1}|^2=\sum_{m=1}^{D-1}\frac{1}{4\sin^2{(\pi m/D})}= (D^2-1)/12,
\end{equation}
using the identity $\sum_{m=1}^{D-1}\mathrm{csc} ^2 (\pi m/D)=(D^2-1)/3$~\cite{wolframcsc}. Also, 
\begin{eqnarray}
\sum_{r\neq 1,c\neq r}|\check{X}_{r,c}|^2= \frac{D+1}{12}(D-1)(D-2),
\end{eqnarray}
using the above identity and the fact that  $\mathrm{csc}^2(\pi (D-1)/D) = \mathrm{csc}^2(\pi /D)$. 
After some simplification the final expression is an upper bound on the decrease in impurity:
\begin{eqnarray}\label{dLf_qft}
dL_{F} 
&\le& -\frac{2}{3}(D+1)\gamma dt L(t),
\end{eqnarray}
which implies that the lower bound on the asymptotic speed-up is 
\begin{equation}
S_{F}\ge\frac{2}{3}(D+1),\label{speedup_lb_qft}
\end{equation}
 as found in Refs.~\cite{ComJac06, ComWisJacOCo10} for a general complementary measurement.

\subsubsection{Upper bound}\label{SEC_2_QFT}
To find the upper bound on the speed-up for the QFT protocol we again uses the method of Ref.~\cite{ComWisJacOCo10}. This instance requires a different fictitious state $\rho_{2}$, again with $L[\rho]=L[\rho_{2}]$, which was termed the binary distribution. The binary distribution is defined as $\rho_{2}={\rm diag} (1-\Delta',\Delta',0,...,0)$ where $1-\Delta'$ is the largest eigenvalue. It is known that $dL$ is most sensitive to permutations of the eigenvalues of $\rho_{2}$ ~\cite{ComWisJacOCo10}.  

Substituting $\rho_{2}$ in to \erf{gen_dpurity2} gives only two terms. Now there is a choice about where to situate the two eigenvalues so that $dL$ is maximized. By calculating the element $|\check{X}_{r,c}|^{2}$, using the QFT, we find that 
\begin{equation}\label{QFT_elem_gen}
|\check{X}_{r,c}|^2=\frac{1}{ 2\left(1-\cos{ \smallfrac{2\pi}{D }(r-c) } \right)}.
\end{equation}
{This allows us to optimisze over the permutations by finding $\max |P_{m}\dg \check{X}_{r,c}P_{m}|^{2}$.
The largest element of the matrix is:  $|\check{X}_{01}|^2 =1/ 2\left(1-\cos{ \smallfrac{2\pi}{D} } \right)$. From this it is clear that $dL$ will be maximized provided the two eigenvalues from $\rho_{2}$ are in succession. The change in the impurity for the binary distribution is}
\begin{eqnarray}\label{dL2_qft}
dL_{\mathrm 2}&=& \frac{-8\gamma dt (2 \lambda_0\lambda_1)}{ 2\left(1-\cos{ \smallfrac{2\pi}{D}  } \right)}  
= \frac{-4\gamma dt}{ \left(1-\cos{ \smallfrac{2\pi}{D}  } \right)}   L_2(t).
\end{eqnarray}
Thus the asymptotic speed-up upper bound is 
\begin{eqnarray}\label{eigenvalueising2}
S_{\mathrm 2} \le \frac{4}{1-\cos{  \smallfrac{2\pi}{D} } } .
\end{eqnarray}
For $D\gg 1$ one finds 
\begin{eqnarray}\label{speed_up_bigD}
S_{\mathrm 2} = \frac{2D^2}{\pi^2}\approx 0.2D^2.
\end{eqnarray}
In \frf{dgen} we depict the conjectured optimal and worst permutations of eigenvalues of a $D$ dimensional state $\rho$. The optimal permutation is $\rho = \mathrm{diag}(\lambda_0,\lambda_{1},\lambda_{3},\lambda_{5},\cdots,\lambda_{6},\lambda_{4},\lambda_{2})$.

\begin{figure}[h!]
\begin{center}
\leavevmode \includegraphics[width=\hsize]{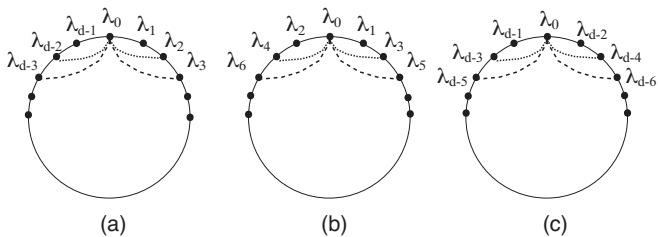}
\caption{a) Weighting factors in \erf{gen_dpurity2} after the feedback has diagonalized and ordered the eigenvalues in $\rho$. b) The conjectured general optimal permutation. c) The conjectured worst permutation. The weights are only shown from the $\lambda_0$ perspective.} \label{dgen}
\end{center}
\end{figure}

In \frf{ass_speed} the numerically calculated asymptotic speed-up as a function of $D$ is plotted. As expected, it is within the lower bound of \erf{speedup_lb_qft} and the upper bound of \erf{eigenvalueising2}. Further it confirms a very nearly quadratic speed-up: the fit shown is $S =0.189D^2+0.109D+0.248$. If, however, we choose to only fit the quadratic term, the fit is
\begin{equation}\label{fitqft}
S=0.19D^2.
\end{equation}
 For $D\ge 10$ this is a good approximation and is very close to \erf{speed_up_bigD}, which was derived for long-times and large $D$. The numerically calculated asymptotic speed-up is only a multiplicative constant away from the ultimate upper bound on all complementary basis purification protocols, the dashed line ($S=0.5D^2 $), \ie\ \erf{speedup_bounds}.

\begin{figure}[h!]
\begin{center}
\leavevmode \includegraphics[width=\hsize]{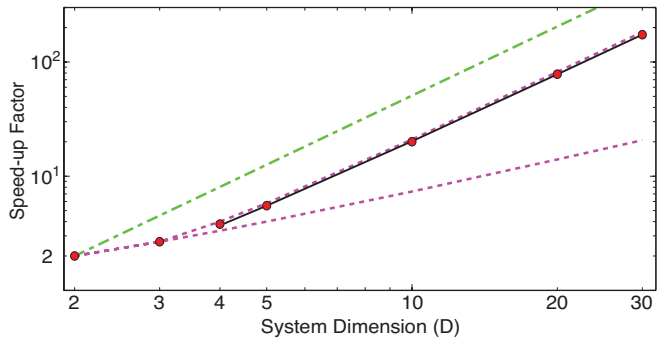}
\caption{The asymptotic speed-up for reaching a given level of purity as a function of the system dimension. The solid line the fit in \erf{fitqft}. The dashed lines are the upper, \erf{eigenvalueising2}, and lower, \erf{speedup_lb_qft}, bounds derived in the text for the QFT protocol. The dot-dashed line is the upper bound on all unbiased basis purification strategies, that is the RHS of \erf{speedup_bounds}.} \label{ass_speed}
\end{center}
\end{figure}

\subsection{Discussion}\label{discusqft}
The results presented in sections \srf{L_Dfb_sec}--\srf{SEC_QFT_bounds} raise a number of questions which we now elucidate further.

Because of symmetries, for dimensions two and three permuting the basis of $\rho$ was unnecessary to maximize $dL$. The intuition is, for maximizing information gain the complementary measurement marginalises the significance of which eigenvalue is the largest. For $D>3$ we must permute the basis of $\rho$, after diagonalization and before applying $T_{\rm QFT}$, to maximize $dL$. The permutations applied do not increase the signal to noise ratio of the measurement unlike the rapid measurement protocol of Ref.~\cite{ComWisJac08}. 
In \srf{nopermsec} and \srf{SEC_2_QFT} we argued that the permutations should be chosen to maximize the product of the two largest eigenvalues in \erf{gen_dpurity2}. While this recipe is operationally sound it lacks physical insight. For example, this recipe does not explain what the mechanism of the purification is and why the permutations can dramatically affect the rate of purification. These questions are addressed in \srf{whycomp}.

The final open question concerns the efficacy of different complementary bases to change the rate of purification. In previous work Lund and Jacobs suggested that not all complementary bases were equally good at entropy reduction \cite{JacLun07}. That is, they claimed that one particular unbiased basis can reduce entropy more quickly than another. If this were true then the $S=0.2 D^{2}$ upper bound we derived from the QFT basis would not necessarily be the true upper bound, which could be closer to $S=0.5D^{2}$ We numerically investigate this question in \srf{dl_all_mub}.

\section{ Why complementarity is not enough}\label{whycomp}

How is it that rotating to an unbiased basis provides any speed-up? Naively following the reasoning applied to qbits in Refs. \cite{FucJac01,DohJacJun01,Jac0303} suggests that making each outcome equally likely (\ie\ maximizing the surprisal~\cite{Tri61}) one maximizes the average amount of information a measurement extracts. From this one might argue that the speed-up observed can be explained by the $D$-dimensional version of this argument. However, it is not clear how this argument would explain why the arrangement of eigenvalues in $\rho$ (permutations) are important to attain the best speed-up. To address this, we examine a phase space representation of the optimal and worst permutations that are schematically depicted in \frf{dgen}. 

Defining a phase space picture for a discrete variable is not trivial. In Ref.~\cite{wig_spin} it was shown that a spin Wigner function $W (\theta,\phi)$ can be defined in terms of Clebsch-Gordan coefficients and spherical harmonic functions. This spin Wigner function is a pseudo-probability distribution on the Bloch sphere, with $\theta$ and $\phi$ the usual Euler angles. The spin Wigner function is a little counter-intuitive; for example, unlike the original Wigner function, $W (x , p)$, for position and momentum the marginal distribution for $\phi$ is not the true phase distribution $P (\phi )$. However, for large $D$ the marginals are a good approximation to the true phase distribution \cite{ComWis05}. We plot the Wigner function using the equal-area projection (described by co-ordinates $\phi$ and $J\cos \theta$).

\begin{figure}[h!]
\begin{center}
\leavevmode \includegraphics[width=\hsize]{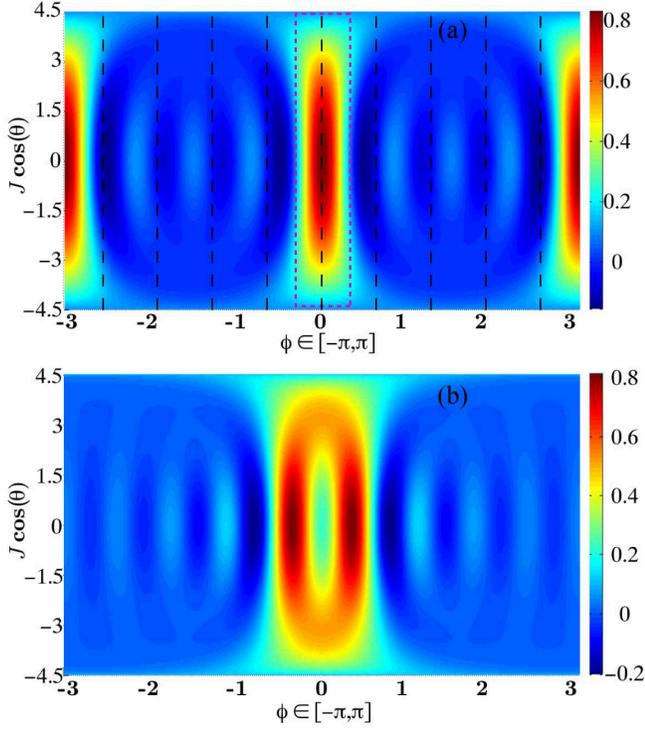}
\caption{An equal area plot of the Wigner function of an equal mixture of two eigenstates after the QFT, $D=10$. a) the worst permutation cf. \frf{dgen} (c). The dashed vertical lines correspond to the locations of the peaks of the $D$ transformed eigenstates. The dotted box around the peak at $\phi=0$ corresponds to the bulk of the Wigner function for this state. b) The conjectured general optimal permutation,  cf. \frf{dgen} (b).} \label{fig13_wig}
\end{center}
\end{figure}

The conjectured best and worst permutations were depicted on a ring in \frf{dgen},  
 explained in terms of the weighting factors $|X_{r,c}|^2$. We now have an intuitive understanding of the angle around this ring as being the phase $\phi$ in the angular representation of spin states. The eigenstates of $\rho$ in the measurement ($J_z$) basis are $\ket{r}$ (Dicke states~\cite{AreCouGil72}). These are transformed 
 by the QFT to the states $\ket{\phi_r}$,  where $\phi_r:=\frac{2\pi}{D}(J-r)$. The states $\ket{\phi_r}=\smallfrac{1}{\sqrt{D}}\sum_{m=-J}^{J}\exp{(-im\phi_{r})}\ket{J,m}$ are   equivalent to the Pegg-Barnett phase states~\cite{PegBar88,PegBar89}. If the QFT were an easy operation in some physical system then our protocol would be  a procedure for rapidly preparing a Pegg-Barnett phase state. 
We note that in some spin systems it should be possible to construct the desired unitary~\cite{MerBreJes09}.

{Consider the long-time-limit state. The worst case, for purification purposes, is when the eigenvalues are equal. Under this two} eigenvalue approximation the worst permutation corresponds to putting the second largest eigenvalue, $\lambda_1$, the maximal distance away from $\lambda_0$ in phase space ($\phi=\pm \pi$). The spin Wigner function for this configuration of the mixture is plotted in \frf{fig13_wig} (a) for $D=10$. 
The optimal permutation is when the two largest eigenvalues are next to each other in phase space; see \frf{fig13_wig} (b). It is now apparent that the schematic diagrams in \frf{dgen} represent a slice through the unwrapped Bloch sphere and the positions of the eigenvalues are their arrangements of the phase states corresponding to the original eigenvalues. This phase space picture suggests  it might be possible to explain why these are the best and worst cases and how the speed-up is generated. To explain these two features we move away from exact calculation of Wigner functions and move to a schematic representation of the Wigner function.
\begin{figure}
\begin{center}
\leavevmode \includegraphics[width=\hsize]{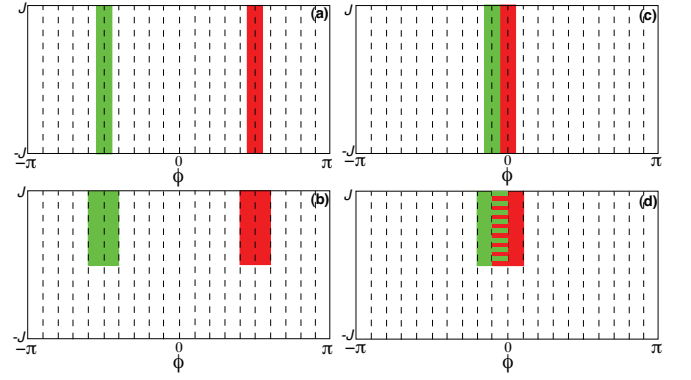}
\caption{Schematic Wigner functions for mixtures of two phase eigenstates for $D=20$. (a) The worst permutation cf. \frf{dgen} (c); we have rotated the {entire sphere by $\pi/2$ about $z$ to simplify} the explanation of the speed-up. b) The worst permutation after a positive measurement result (modulo $\pi/2$). The total area is constant in figures (a) and (b), and is equal to $4\pi$. (c) The optimal permutation cf. \frf{dgen} (b). (d) The optimal permutation after a positive measurement result. The total area is reduced in figure (d) to $3\pi$, which gives rise to a purifying effect (the purity is inversely proportional the area underneath the Wigner function). The striped region $\phi \in (-\pi/D,0]$ denotes that the red and green rectangles are overlapping in this region. From these figures it is apparent that the circles in \frf{dgen} can be thought of as showing $\phi$.} \label{fig14}
\end{center}
\end{figure}

In \frf{fig14} (a) we represent schematically the bulk of the Wigner function by a rectangle of width $2\pi/D$ (corresponding to the $\phi$ coordinate) and height $D-1=2J$ (which corresponds to the $J\cos \theta$ coordinate). In what follows one may loosely think of the rectangle as representing the uncertainties in an observer's knowledge about two conjugate variables; according to the \hei uncertainty relation (HUR) the area of the rectangle must be constant for pure states. {Now consider the effect of a weak measurement of $J_{z}$. In our protocol this is} of infinitesimal duration, but here we exaggerate the effect to illustrate our point. In \frf{fig14} (b) we have taken the result of the measurement to be positive. Because the result was positive, the Wigner function does not have much support on the lower part of the plot. As the positive result contained information about the distribution of $j$, the uncertainty in this variable is reduced. In keeping with the HUR, the conjugate variable ($\phi $) suffers an increase in variance. For Wigner functions the purity is proportional  to the inverse of area under the function \cite{Schl01}. Here there has been no change in the total area and hence no change in the total purity. {Because this is only a heuristic for understanding protocol the previous statement is not entirely true; the final paragraph of this section will explain the actual result.}


In \frf{fig14} (c) the optimal permutation in the QFT basis is depicted. The total area of the two rectangles before the measurement is $\sim 4\pi$. After a positive measurement result the phase distributions for the two eigenvalues significantly overlaps. The total area is now $\sim 3\pi$; this reduction in area leads to an increase in purity; see \frf{fig14} (d). Now the intuitive understanding for the speed-up and the permutation sensitivity is apparent. The permutations are important so that the large eigenvalues in $\rho$ may bleed into each other after a measurement. The bleeding is due to measurement backaction in the variable conjugate to $J$. It is the reduction in area this bleeding effects that causes the purification; but it only works if the largest two eigenvalues are adjacent in phase space. Although this picture is crude it captures the essence of the protocol. 

So far we have given an intuitive explanation of the mechanism of the purification. Now we provide an intuitive explanation for the speed-up. Consider the integrated measurement result 
\begin{equation}
\int _{0}^{t}dR(t)= \int _{0}^{t}\sqrt{4\gamma} X(s)ds+\int _{0}^{t}dW(s),
\end{equation}
this expression contains a term corresponding to the signal of interest (the first term) and a term representing the noise (the second term). We may define the signal to noise ratio as ${\rm SNR}={\rm signal}^{2}/{\rm noise}^{2}$. In this case we have ${\rm signal}^{2}= 4\gamma \Delta t^{2} \an{X^{2}}$ and ${\rm noise}^{2} = \an{(\Delta W(t))^2} = \Delta t$. Thus the ratio becomes ${\rm SNR} = 4\gamma \Delta t 
\an{X^{2}}$. Now we wish to estimate the time taken to evolve from the Wigner functions in the first row of \frf{fig14} to those in the second row. Note that an observer gains one bit of information about $J_{z}$ between the first and second rows. From the Shannon--Hartley theorem, ${\rm bits}= \log_{2}(1+ S/N)$, we infer that this implies ${\rm SNR} = 1$. Now we solve for time and take $X=J_{z}$ consequently $ \an{J_{z}^{2}}= (D^{2}-1)/12$ (see appendix C in Ref.~\cite{ComWisJacOCo10}) to find
\begin{equation}
\Delta t = \frac{3}{\gamma (D^{2}-1)}.
\end{equation}
Recall that the purity is approximately inversely proportional to the phase space area. The initial purity in \frf{fig14} (c) is $P_{i}\propto 1/4 \pi$ and we know the state is an equal mixture of two eigenstates which means the purity equals one half. This allows us to determine the proportionality constant  to be $2 \pi$, so $P_{i}=\half$. The final purity is thus $P_{f}=2/3$ which means the change in purity is $\Delta P = 1/6$. Consequently 
\begin{equation}
\frac{1}{P}\frac{\Delta P}{\gamma \Delta t}=\frac{1}{2}\frac{(1/6)}{3/(D^{2}-1)}=\frac{1}{9}(D^{2}-1),\label{naive_dp}
\end{equation}
This can be compared to \erf{speed_up_bigD} that is
\begin{equation}
\frac{1}{L}\frac{dL}{\gamma dt}=-\frac{2D^{2}}{\pi^2}\approx -\frac{1}{5}D^{2}.\label{exact_dl}
\end{equation}
{Considering the crudeness of the arguments we have employed, there is surprisingly good } agreement between \erf{naive_dp} and \erf{exact_dl}.  This gives weight to the intuition that the effect of purification for the QFT protocol comes from measurement back-action and the role of the permutations is to maximize the back-action by placing the eigenvalues close in phase space.

{We now return} to the worst permutations for purification.  All one can conclude from the equi-likelhood of all outcomes is that there should be a speed-up of at least two (in a discrete outcome measurement model, which is equivalent to our current measurement model). Let us confirm this intuition now. Returning to the two eigenvalue approximation, consider the worst permutation. It is possible to determine the upper bound on speed-up resulting from this permutation by substituting $r-c= D/2$ into \erf{QFT_elem_gen}. We find that $S=2$, which confirms the intuition above. This is why complementarity is not enough to guarantee maximal information extraction.


\section{Are all complementary bases equally good for entropy reduction?}\label{dl_all_mub}
We now address the question raised by Jacobs and Lund~\cite{JacLun07}: are all unbiased bases equally good for rapid purification?   Their answer was no. However at the time of their analysis the role of permutations was not clearly understood. Thus it is worthwhile to re-examine this question. Due to rotational symmetry {of the unbiased bases and the permutational symmetry of density operator} the answer in the cases $D=2$ and $3$ is yes. However for $D=4$ it is easy to find a counter-example to this trend. For example measuring in any of the four {\em mutually unbiased bases} (MUBS) ~\cite{mubs} gives
\begin{eqnarray}\label{d4weightsmubs}
\nn dL&=&\!- 8\gamma dt \sum_{r,c} |\breve{X}_{r,c}|^2\lambda_r\lambda_c\\
\nn &=&\!-8\gamma dt \left[2\lambda_0\lambda_1+\half ( \lambda_1\lambda_2+ \lambda_0\lambda_3) + 2\lambda_2\lambda_3 \right.\\
&&\phantom{\!-8\gamma dt [}+\left .0(\lambda_{0}\lambda_{2}+\lambda_{1}\lambda_{3})\right ],
\end{eqnarray}
here $\breve{X} = M_{i}^\dag J_{z} M_{i}$ where $M_{i}$ is a transformation to one of the MUBS. (Explicit expressions of the five MUBS when $D=4$ are given in Ref. \cite{mubs} for example. For convenience we have reproduced them in a footnote \footnote{ $M_{0} = [[1,0,0,0]',[0,1,0,0]',[0,0,1,0]',[0,0,0,1]']$, $M_{1} = [[1,1,1,1]',[1,1,-1,-1]',[1,-1,-1,1]',[1,-1,1,-1]']/2$,
 $M_{2} = [[1,-1,-i,-i]',[1,-1,i,i]',[1,1,i,-i]',[1,1,-i,i]']/2$,
 $M_3 = [[1,-i,1,i]',[1,i,1,-i]',[1,-i,-1,-i]',[1,i,-1,i]']/2$,
 $M_4 = [[1,-i,-i,-1]',[1,-i,i,1]',[1,i,i,-1]',[1,i,-i,1]']/2$.}.) As before, we may use $\rho_{2}$ to obtain a lower bound on $dL$ (and hence an upper bound on the speed-up). Doing so gives $dL = -8\gamma dt L(t)$ which implies a speed-up of $S=8$. This saturates the upper bound on the speed-up found in \erf{speedup_bounds}. The factor of two improvement over the QFT can be understood from the difference in weights of the largest terms $\lambda_{0}\lambda_{1}$ between \erf{d4weights} and \erf{d4weightsmubs}. 
 
 \begin{figure}[h!]
\begin{center}
\leavevmode \includegraphics[width=\hsize]{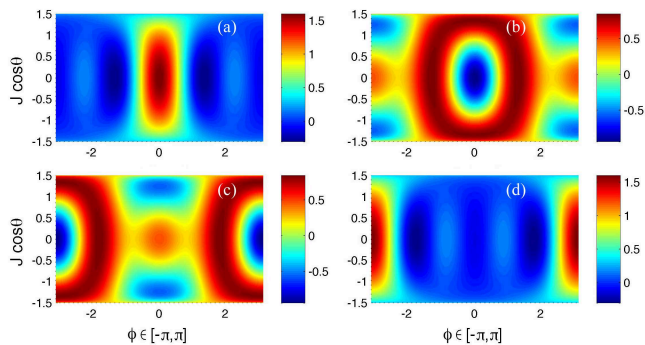}
\caption{An equal area plots of the Wigner function resulting from transforming four different states to the $M_{1}$ unbiased basis ($D=4$). (a) the transform of $\rho_{0}={\rm diag}(1,0,0,0)$ (b) $\rho_{0}={\rm diag}(0,1,0,0)$ (c) $\rho_{0}={\rm diag}(0,0,1,0)$ and (d) $\rho_{0}={\rm diag}(0,0,0,1)$. Unlike the unbiased basis generated by the QFT not all of the resulting states are like the Pegg-Barnett phase states; see for example figures (b) and (c).} \label{mubeig}
\end{center}
\end{figure}

To get more intuition about the purification process using the MUBs transform we look at the Wigner function of the {four states in one of the four MUBs,  which for convenience we denote $\{\ket{0},\ket{1},\ket{2},\ket{3}\}$, in \frf{mubeig}.} The states plotted in \frf{mubeig} (a) and (d) look like the states in the QFT basis (Pegg-Barnett phase states). The Wigner functions plotted in \frf{mubeig} (b) and (c) are however quite different --- they contain a hole which is more negative than the dips in \frf{mubeig} (a) and (d). The states plotted in \frf{mubeig} (b) and (c) can be said to be highly non classical because of this.

We now compare the MUB transform to the QFT using the Wigner function, in an attempt to get an intuitive explanation of the advantage of the MUB transform. To make the analysis simple we restrict to states which have two large eigenvalues only (as we did above). From \erf{d4weightsmubs} we see that permuting these states in the logical basis so that they correspond to $\ket{0}$ and $\ket{1}$ (or $\ket{2}$ and $\ket{3}$) is the optimal thing to do. Recall that permuting the two largest eigenvalues to $\ket{0}$ and $\ket{1}$ was also optimal for the QFT protocol; see \frf{qftmubcf} (a). There we found that doing so created two peaks close in phase space, so the purification could be explained by measurement backaction. In \frf{qftmubcf} (c) the Wigner function of the optimally permuted state for the MUB transform is depicted [it is an equal mixture of \frf{mubeig} (a)and (b)]. Here we can no longer attribute the purification mechanism to backaction. In fact it seems as though the purification effect comes from distinguishing the two peaks in the $J_{z}$ distribution. The worst QFT permutation also corresponds to a poor MUB transform as evidenced in \frf{qftmubcf} (b) and (d).

\begin{figure}[h!]
\begin{center}
\leavevmode \includegraphics[width=\hsize]{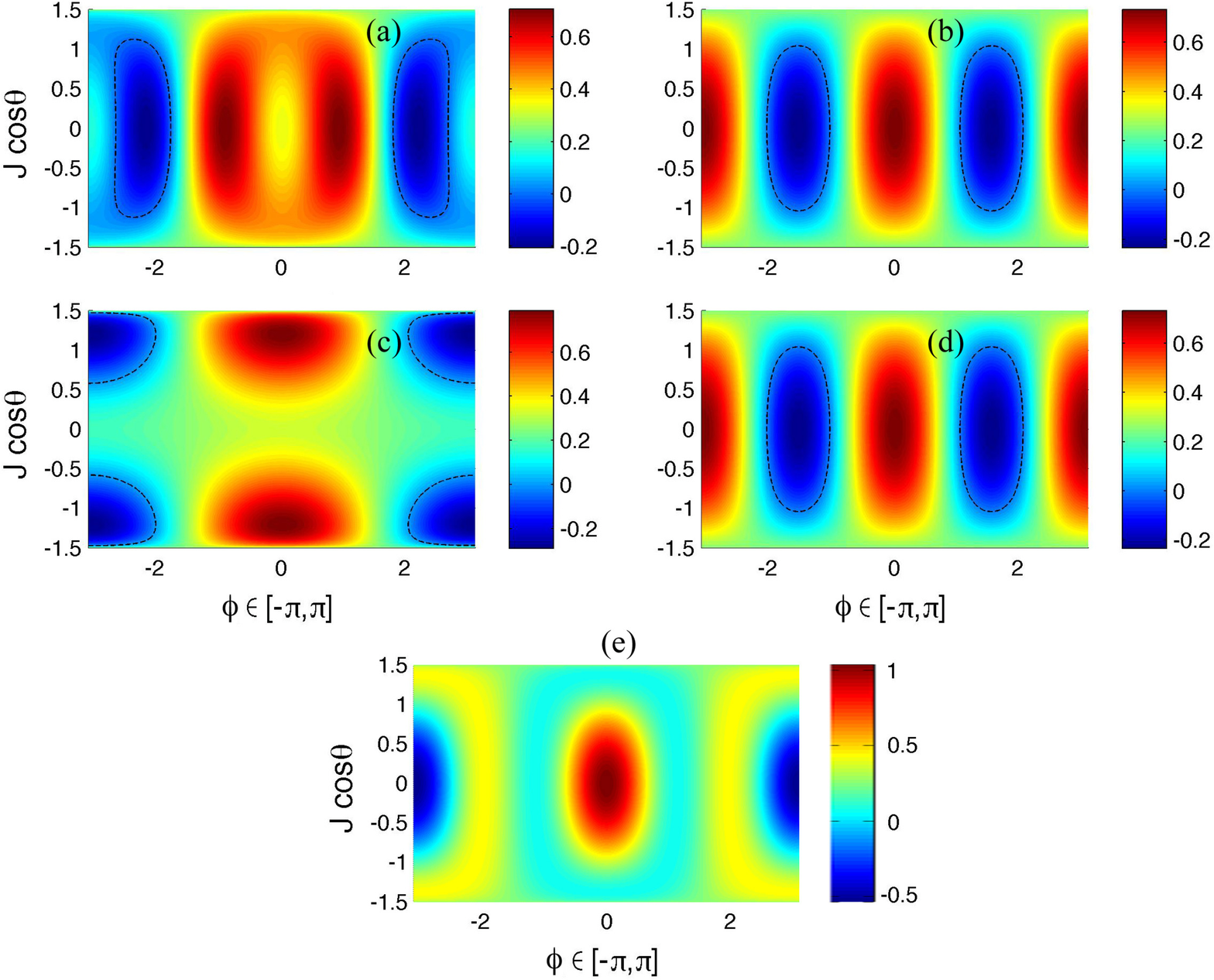}
\caption{A Wigner function comparison between the QFT (top row) and MUB transform (bottom row) for $D=4$. We plot an equal mixture of two eigenstates with the optimal permutation (first column) and a poor permutation (second column) for each transform. In figures (a)--(d) the black dashed line is the zero contour. (a) the optimally permuted state for the QFT -- $\rho ={\rm diag}(1,1,0,0)/2 $; (b) the worst permutation for the QFT -- $\rho ={\rm diag}(1,0,1,0)/2 $; (c) the optimally permuted state for the MUB transform -- $\rho ={\rm diag}(1,1,0,0)/2 $; (d) a poor permutation for the MUB transform -- $\rho ={\rm diag}(1,0,0,1)/2 $. Note that cases (b) and (d) are actually identical. 
Unlike the QFT case, the mixture of optimally permuted MUB states has two peaks which could be distinguished by a measurement of $J_{z}$.   (e) The worst permutation \ie\ $\rho ={\rm diag}(1,0,1,0)/2 $ after the transform to the $M_{1}$ unbiased basis, which results in {\em no} purification.  
} \label{qftmubcf}
\end{center}
\end{figure}

 Interestingly there is a permutation that is worse for the MUB transform than the permutation depicted in \frf{qftmubcf} (d). It is $\rho ={\rm diag}(1,0,1,0)/2 $; the transformed state of this permutation is depicted in \frf{qftmubcf} (e). This permutation results in {no} purification at all!  We can see this by substituting $\lambda_{0}=\half$ and $\lambda_{2}=\half$ into \erf{d4weightsmubs} to get $dL=0$. 
This complete lack of purification may have application in state stabilisation of states with arbitrary purity 
\cite{JacLun07}. 

We would like to know if the upperbound on the speed-up is saturated in all dimensions. To answer this question we resort to a numerical search for $D\in[2,10]$. We do not claim that our search is exhaustive. Recall that the stochastic simulations show that speed-up predicted by $\rho_{2}$ is close to the achievable amount. It seems reasonable to assume this to be true in other complementary bases. This greatly simplifies the analysis by making the optimisation of permutations superfluous. Because of this we may numerically search for over all unbiased bases for the basis which has the largest element $|\breve{X}_{r,c}|^2$. Converting the element $|\breve{X}_{r,c}|^2$ to a speed-up gives the following trend. The speed-up for even $D$ attains  the upperbound, \ie\ $D^{2}/2$, and the speed-up for odd $D$ equals $(D-1)^{2}/2$, as seen in \frf{smaxscale}. In either case this is much larger than the speed-up of the QFT protocol which was $S=0.2D^{2}$. For $D\gg 1$ it is reasonable to believe that the achievable speed-up does indeed scale like the predicted $S=0.5D^{2}$ \cite{ComWisJacOCo10}.

\begin{figure}[h!]
\begin{center}
\leavevmode \includegraphics[width=\hsize]{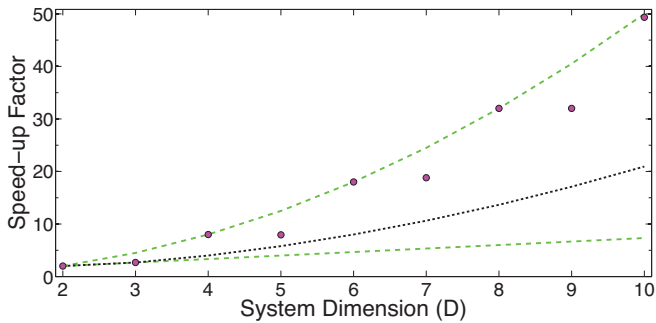}
\caption{The asymptotic speed-up for reaching a given level of purity as a function of the system dimension. The dashed lines are the upper bound and lower bounds on all unbiased basis purification strategies, \ie\ \erf{speedup_bounds}. The dotted line is the maximum achievable speed-up for the QFT protocol \ie\ \erf{eigenvalueising2}. The solid circles are the results of a numerical search over unbiased bases to maximize the asympototic speed-up. See the text for more details.} \label{smaxscale}
\end{center}
\end{figure}


\section{Information acquisition in a register of qbits and its relationship to complementarity}\label{qftreg}


Finally, we consider complementary measurements of a register of $n$ qbits, where each qbit is independently and weakly (or continuously) measured, as introduced in Ref.~\cite{ComWisJac08}. Instead of one observable $X$, there are now $n$, given by $X^{(r)}= I^{(1)}\otimes I^{(2)}\otimes \ldots \sigma_z^{(r)}\ldots \otimes I^{(n)} $, where $r$ labels the $r$th qbit. The SME describing such a measurement is 
\begin{equation}\label{sme_reg}
  d\rho = \sum _r 2\kappa \, dt\,\D{X^{(r)}}\rho+\sqrt{2\kappa}\,dW^{(r)}\Hc{X^{(r)}}\rho.
\end{equation}
The combined state of the $n$ qbits exists in a $D=2^n$-dimensional Hilbert space. 

\subsection{Commuting measurements analysis}\label{regnfb}
Here we will not analyse the commuting measurements of a register of qbits with the same detail as we did for the qdit in \srf{CommMeas}. Instead we rely on a result from Ref.~\cite{ComWisJacOCo10} where an expression for the evolution of the average impurity of a register by undergoing a continuous commuting measurement was found. In the long time (LT) limit the impurity is
\begin{equation}\label{nfbreg}
   \expt{ L^{(n)}(t)}_{\rm LT} =  \frac{n\pi e^{-4\kappa t}}{8\sqrt{\pi\kappa t }}.
\end{equation}
We will use this expression in this section to calculate the speed-up, so only the asymptotic scaling is important: $\expt{ L^{(n)}(t)}\sim e^{-4\kappa t}$. 

\subsection{Complementary measurements}\label{SEC_reg_lb}
It was shown in Refs.~\cite{ComWisJacOCo10} that the change in impurity for a register of qbits monitored in an unbiased basis is
\begin{eqnarray}
 dL&=&- 8\kappa \,dt \sum_{r=1}^n \tr{ \check{X}^{(r)}\rho \check{X}^{(r)}\rho}\\
 &=&- 8\kappa \,dt \sum_{r=1}^n\sum_{i,j=0}^{(D-1)} |\check{X}_{i,j}^{(r)}|^2\lambda_i\lambda_j \label{dl_reg_xij}.
\end{eqnarray}
Here we define the complementary observable to be $\check{X}^{(r,m)}= P_m T X ^{(r)} T\dg P_m \dg$. As before, the $T$'s are conditional unitaries that introduce the unbiasedness (between $\rho$ and $X^{(r)}$), and the $P_m$'s are the permutation operators.

It was found that the upper bound on the impurity was
\begin{eqnarray}\label{reg_ub}
L^{(n)}_{\rm UB}(t)&=&  \exp{(-8\kappa nt/[D-1])}L(0),
\end{eqnarray}
while the lower bound was
\begin{eqnarray}\label{L_reg_lb}
L^{(n)}_{\rm LB}(t)&=&\exp{(-8\kappa n  t)} L(0).
\end{eqnarray}
From these equations we can infer the following bounds on the asymptotic speed-up factor 
\begin{eqnarray}\label{regspeed}
 \frac{2n}{D-1}\le S\le 2n.
\end{eqnarray}
For all $n\ge 3$ the lower bound on the speed-up becomes less than unity. For large $n$ the slow-down implicit in the lower bound is $\sim n2^{-n+1}$. (The interested reader should also see the related study of Hill and Ralph~\cite{HilRal0705}.)

\subsection{Complementarity via the quantum Fourier transform}\label{SEC_QFT_reg}
From \erf{regspeed} it is not clear if measurements in a complementary basis provide any benefit in information extraction. In this subsection we present some progress towards answering this question.

We begin our analysis by considering the speed-up when $n=2$. The bounds on the speed-up are $4/3\le S \le 4$. In order to see if the upper bound is achievable we perform stochastic simulations. In \frf{regsp} we numerically determine the advantage of the QFT feedback protocol in a register of two qbits. The permutation used for simulating the evolution was $\rho = (\lambda_0,\lambda_4,\lambda_3,\lambda_1)$.

\begin{figure}[h!]
\begin{center}
\leavevmode \includegraphics[width=\hsize]{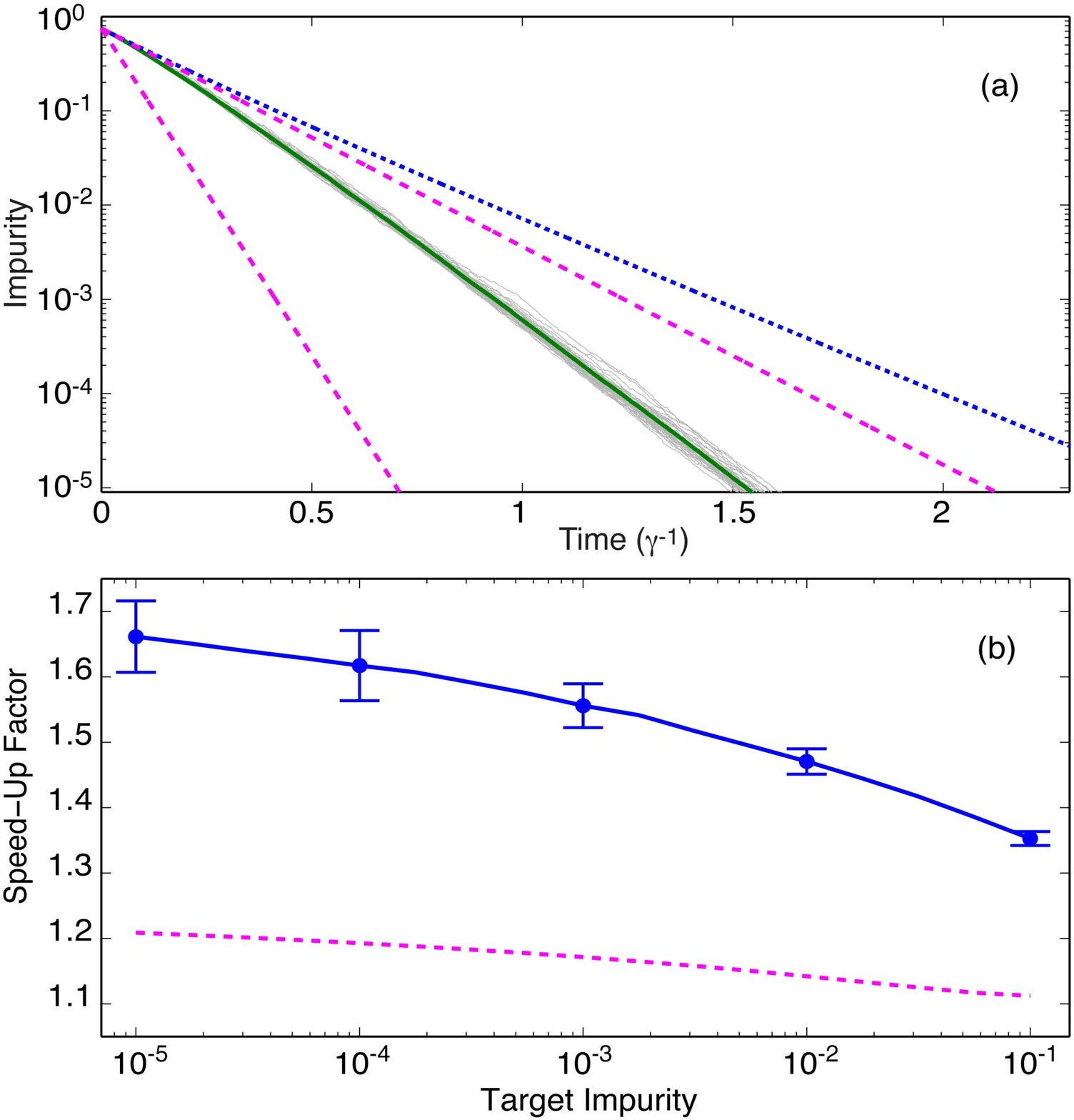}
\caption{ (a) Impurity for a two qbit register: dotted line -- a continuous measurement; solid green line -- average impurity for feedback; solid grey lines -- individual trajectories grey lines; dashed magenta lines - upper [\erf{reg_ub}]  and lower [\erf{L_reg_lb}] bounds. (b) Speed-up solid line; lower bound dashed line; upper bound not shown as it well above the numerical result. The ensemble size is $20$ and feedback was applied at times $\delta t = 1\times 10^{-3}\gamma^{-1}$.} \label{regsp}
\end{center}
\end{figure}

In \frf{regsp} (a) the QFT protocol clearly does not saturate the bound on $\expt{L}$ given in \erf{L_reg_lb}. One may randomly permute the eigenvalue arrangement at times $\delta t$ before applying the QFT feedback to determistically achieve the lower bound of \erf{regspeed}. From \frf{regsp} (b)  it seems as though the speed-up is asymptoting towards 2 rather than the upper bound specified by \erf{regspeed}, i.e. 4. For comparison, we note that the speed-up found for the locally optimal rapid measurement protocol in Ref.~\cite{ComWisJac08}  was $S_{\rm RM}\approx1.4$  in the long-time limit. 

The structure of unbiased bases for a register of qbits is quite complicated \cite{LawBruZei02, RomKilSan05, KilSanGui05}. Even for the QFT the optimal permutation is not obvious for $n>2$. It is for this reason we plot the two eigenvalue approximation to the asymptotic speed-up for the QFT in \frf{reg_asym} as a function of $n$. The values were obtained by finding the largest term in \erf{dl_reg_xij}. This term will be denoted by $X_{\rm max}:=\max_{i,j}\sum_r|X_{i,j}^{(r)}|$. The feedback places the two eigenvalues at $i_{\rm max}$ and $j_{\rm max}$. Thus $X_{\rm max}$ is proportional to the speed-up. Curiously the speed-up, 
\begin{equation}\label{speed_reg_qft}
S\approx 2,
\end{equation}
appears to be independent of the size of the register. It is not clear if this is true for any unbiased basis in a register. Our arguments from \srf{dl_all_mub} indicate it might be a multiplicative constant higher.

\begin{figure}[h!]
\begin{center}
\includegraphics[width=\hsize]{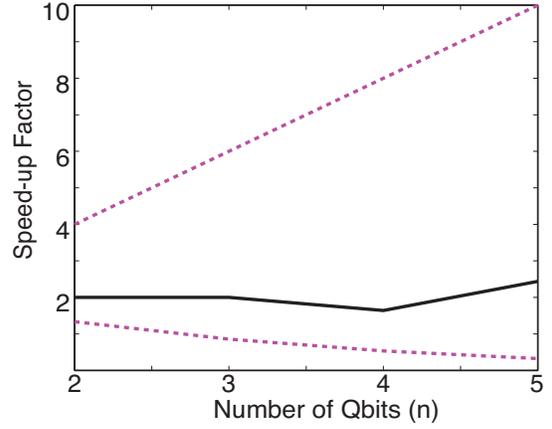}
\caption{Bounds on the asymptotic speed-up for a register of qbits. The dashed lines are the upper and lower bounds derived in the main text \erf{regspeed}. The solid line is the largest element of $X_{\rm max}=\max_{i,j}\sum_r|X_{i,j}^{(r)}|$. The element $X_{\rm max}$ is conjectured to be a good indicator for maximum achievable speed-up with the QFT in a register of qbits.} \label{reg_asym}
\end{center}
\end{figure}




\section{Discussion}\label{discuss}

Prior to this work it has been shown that it is possible to speed-up the extraction of information from a quantum system using a continuous complementary measurement \cite{Jac0303, ComJac06, ComWisJacOCo10}. In this paper we have given an explicit method for constructing such a method, using the QFT, that achieves a speed-up of $S = 0.2D^{2}$ for a qdit. This explicit construction allowed us to show that complementary between the state and measurement observable is not enough to guarantee maximal information extraction. Choosing the right permutation of eigenstates, {before Fourier transforming the state -- see \frf{dgen},} is needed to guarantee maximal information extraction, which is contrary to what one might expect from classical information theory \cite{Tri61}. {The effect of the permutation can change the speed-up in information extraction from its minimal value $2$ to $0.2D^{2}$.} We also argued that in the case of the qdit,  the maximum possible speed-up, predicted in Ref.~\cite{ComWisJacOCo10}, $S\sim D^{2}/2$ should be achievable.  One interesting upshot of our investigation (see \srf{dl_all_mub}) is an explicit example of using an unbiased basis for state stabilisation \cite{JacLun07}.

In this paper  we demonstrated that measurement backaction is fundamental to the purification process for the QFT protocol. We did not, unfortunately, find an information theoretic explanation of the measurement process as was found in Ref.~\cite{KurKasCze10}. It is not clear, at present, that the bounds presented in \srf{SEC_QFT_bounds} could be obtained from more elementary reasoning about uncertainty relations \cite{MasSpi08}. 

In related work, Shabani and Jacobs~\cite{ShaJac08,JacSha08} found the globally optimal (in time) protocol for $D=3$ and the locally optimal protocol for all dimensions for reduction of a quantity related to impurity. Their calculated bounds on the speed-up were $2(D-1)\le S_{\rm SJ} \le 2(D-1)^{2}$. Naively comparing their bounds to ours suggest that an unbiassed protocol is not optimal. However it is not yet clear if their upper bound is achievable for $D>4$. 

It is possible to perform a simple calculation, independent of the one presented in Ref.~\cite{ShaJac08},  for the absolute upper bound on the speed-up. First we calculate a lower bound on ${\rm E}[dL]$:
\begin{eqnarray}
\nn{\rm E}[dL]\!\!&=&\!\! - 8\gamma dt \Big \{ \tr{\rho \breve{X}\rho \breve{X}}-2\tr{\breve{X}\rho}\tr{\breve{X}\rho^2}\\&&\phantom{- 8\gamma dt} \,+\tr{\rho \breve{X}}^2\tr{\rho^2}\Big \},
\end{eqnarray}
by using the trace inequality $\tr{ABAB}\le \tr{A^{2}B^{2}}$, for $A,B$ Hermitan. The breve on $X$ denotes the unitary transformation $\breve{X}= U\dg X U$, where $U$ is {\em any} unitary. Note that all but the first term, $\tr{X\rho X\rho}$, are are $O(1-L^{2})$ [or $O(L^{2})$]. Consequently the first term dominates ${\rm E}[dL]$ as it is $O(L)$. Next we make a two eigenvalue approximation to $\rho$. We find
\begin{eqnarray}
{\rm E}[dL_{2}]&=& - 8\gamma dt \tr{\rho_{2} \breve{X}\rho_{2} \breve{X}}\\
&\le&-8\gamma dt \tr{X^{2}U\rho_{2}^{2}U\dg}\le2(D-1)^{2}.
\end{eqnarray}
This implies an upper bound on the speed-up of $S= 2(D-1)^{2}$, as found by Shabani and Jacobs \cite{ShaJac08}. The results of a numerical search, like the one performed in \frf{smaxscale}, indicates that this bound is indeed achievable. Furthermore we can confidently say that the speed-up of the time-optimal control strategy for impurity reduction is bounded above by $S= 2(D-1)^{2}$. This is justified by making a two eigenvalue approximation to $\rho_{2}$ and then using the proof that Jacobs' protocol \cite{Jac0303} is optimal \cite{WisBou08,ShaJac08,JacSha08,BelNegMol09}. 

It is interesting that the rapid measurement protocol (considered in Refs.~\cite{ComWisJac08,ComWisSco10}) and the complementary measurement protocols both afford  {\em at most} a speed-up $O(D^{2})$. They work in very different ways. The rapid measurement protocol is essentially classical in nature. It uses operations that ensure the state and measurement commute at all times. {In particular the operations are permutations of the state in the measurement basis. The phase space picture for the rapid measurement protocol, in the long time limit (i.e. under the two eigenvalue approximation), would be two $J_{z}$ eigenstates (Dicke states) placed at $\pm J$. As the measurements are in the $z$ basis,} the mechanism for the rapid measurement speed-up is an increase in signal to noise ratio {in the measurement record}. In this article we have shown that the complementary measurement protocol is essentially quantum mechanical in nature. The mechanism for the purification and speed-up is measurement backaction (at least for the QFT protocol). The two approaches also have different advantages. Rapid measurement has the advantage that it enables one to obtain information about the initial state. The complementary protocol presented here has the advantage that it provides a nearly deterministic improvement in information gathering of the current system state. In future work we will explore if it is possible to obtain information about the initial state for a continuous complementary measurement.

{Finally we will speak to the practicality of implementing QFT based protocols. In recent work we showed how to analyse imperfections in a purification protocol using a feedback master equation \cite{ComWis11a} so we will not discuss such issues further here. Instead we will focus on the implementation the QFT. For qbits
the QFT is a fundamental quantum logic gate known as the Hadamard gate \cite{NieChu00}, and is typically easy to implement. 

For $D>2$ implementing the QFT is more difficult, as it involves multiple logic gates (for $n$-qbit systems) \cite{Chi05} or multi-level coherent operations (for a qudit). Earlier we mentioned that in some atomic spin systems it should be possible to construct the desired QFT unitary~\cite{MerBreJes09}. We think, however, that solid state systems are the most likely candidate for which it would be useful to implement the ideas presented in this article. This is because in solid state systems the measurement strength $\gamma$ is typically much smaller (by an order of magnitude or more) than the maximum control strength $|\alpha|$ (e.g. in Ref.~\cite{Pet05} $1/\gamma$ is on the order of microseconds and $1/|\alpha|$ is of order nanoseconds). Consequently it is possible to imagine applying feedback continuously through the measurement process so that the eigenbasis of the state and observable are QFT pairs. Alternatively, if the maximum control strength is large enough, the feedback could be applied impulsively at discrete times, which can gives results surprisingly close to the continuous version (as noted in \frf{d4speed} and Ref.~\cite{ComWisSco10}). Further the exciting results in the field of superconducting qdits ($D\in [3,5]$) where measurement and full unitary control has been demonstrated for a number of systems \cite{Bia10,Nee09} lead us to speculate that our proposals could be experimentally tested within the next decade.}

{\em Acknowledgements:} The authors acknowledge helpful discussions with Kurt Jacobs, Andy Chia, Paulo E. M. F. Mendon\c{c}a, Andrew Scott, Alireza Shabani, Alexandre B. Tacla {and Carlton M. Caves}. This research was supported by the Australian Research Council Centre of Excellence CE110001029 \& CE110001027.  JC also acknowledges support from National Science Foundation Grant No. PHY-0903953 and Office of Naval Research Grant No. N00014-11-1-008.

\bibliographystyle{apsrev}

\end{document}